\documentclass[aps,prd,10pt,twocolumn,showpacs,amsmath,amssymb,nofootinbib,superscriptaddress]{revtex4-2}
\usepackage{graphicx}  
\usepackage{color}
\usepackage{times}
\usepackage{float}
\usepackage{afterpage}
\usepackage[utf8]{inputenc}
\usepackage{bm}
\usepackage{ulem}
\usepackage{multirow}
\usepackage{makecell}
\usepackage{url}
\usepackage{natbib}
\usepackage{mathrsfs}
\usepackage{physics}
\usepackage{comment}
\usepackage[table,xcdraw]{xcolor}
\usepackage{booktabs,makecell}
\usepackage{amsmath}
\usepackage{pifont}
\usepackage[colorlinks=true,citecolor=blue,urlcolor=blue,linkcolor=blue]{hyperref}
\usepackage{microtype}
\usepackage[none]{hyphenat}
\begin{document}

% Title and author information
%\title{Structure, Stability, and Oscillations of Neutron Stars with Gravitationally Bound Dark Matter}
\title{Asteroseismology and Universal Relations in Neutron Stars with Gravitationally Bound Dark Matter}

% Authors and affiliations
\author{Ankit Kumar}
\email{ankit.k@iopb.res.in}
\affiliation{Department of Mathematics and Physics, Kochi University, Kochi, 780-8520, Japan}

\author{Hajime Sotani}
\affiliation{Department of Mathematics and Physics, Kochi University, Kochi, 780-8520, Japan}
\affiliation{RIKEN Center for Interdisciplinary Theoretical and Mathematical Sciences (iTHEMS), RIKEN, Wako 351-0198, Japan}
\affiliation{Theoretical Astrophysics, IAAT, University of T\"{u}bingen, 72076 T\"{u}bingen, Germany}

\date{\today}

%%%%%%%%%%%%%%%%%%%%%%%%%%%%%
% Abstract
\begin{abstract}
We investigate the structural, dynamical, and oscillatory properties of neutron stars admixed with dark matter, modeled via a single-fluid formalism where dark matter interacts with nuclear matter through an effective Higgs-portal coupling. Employing three relativistic mean-field nuclear matter equations of state—IOPB-I, BigApple, and NL3— we incorporate a physically motivated dark matter number density profile that scales with baryon density and is controlled by two parameters: a scaling factor $\alpha M_\chi$ ($M_{\chi}$ being the mass of dark matter particle) and a steepness index $\beta$. We construct equilibrium configurations and analyze their stability via radial oscillations, finding that dark matter-induced gravitational compression lowers the maximum mass and alters the radial mode spectrum in a nontrivial, $\beta$-dependent fashion. We also compute the frequencies of non-radial fluid oscillations under the relativistic Cowling approximation and analyze the persistence of universal relations in the presence of dark matter. While deviations appear under extreme configurations, the overall structure of these relations remains robust. Our findings offer a consistent framework to probe dark matter effects on neutron star dynamics across a range of realistic models.
\end{abstract}
%%%%%%%%%%%%%%%%%%%%%%%%%%%%%

% PACS numbers
%\pacs{Insert PACS numbers here}

\maketitle

% Introduction
%%%%%%%%%%%%%%%%%%%%%%%%%%%%%
\section{Introduction}
\label{sec:1}
%%%%%%%%%%%%%%%%%%%%%%%%%%%%%
Neutron stars (NSs) are astrophysical objects of extreme density and gravity, making them ideal for studying the physics of dense matter, gravitational interactions, and fundamental particle interactions. Their composition is governed by the equation of state (EOS), which relates pressure to energy density and dictates key stellar properties such as mass, radius, and oscillation frequencies~\cite{annurev:/content/journals/10.1146/annurev-astro-081915-023322}. However, the nature of matter at supra-nuclear densities remains uncertain, prompting continued investigations through gravitational waves, X-ray observations, and theoretical modeling~\cite{Bogdanov_2019, Bogdanov_2021, PhysRevLett.119.161101}. Beyond the conventional nuclear matter, the potential presence of dark matter (DM) within these compact objects introduces additional complexity to their internal structure and dynamics~\cite{PhysRevD.40.3221, PhysRevD.92.063526, Shawqi_2024}. The impact of DM on NSs depends on whether DM forms a core, a halo, or is homogeneously distributed within the star \cite{Giangrandi_2023, PhysRevD.111.043050, 10.1093/mnras/stae337, Konstantinou_2024}. Theoretical models for DM admixed NSs often fall into two main categories: (1) Two-fluid models, where nuclear and dark matter components are treated as separate fluids interacting only gravitationally \cite{PhysRevD.102.063028, 10.1093/mnras/stad3658, PhysRevD.110.023013, kumar2025mcc}; (2) Single-fluid models, where DM is incorporated directly into the nuclear EOS, allowing for interaction effects such as DM-baryon interactions \cite{PhysRevD.96.083004, PhysRevD.104.063028, PhysRevD.106.043010, 10.1093/mnras/staa1435}.  The presence of DM in NSs has significant implications for astrophysical observations, particularly in the context of mass-radius constraints from pulsars and X-ray binaries. DM self-interactions and particle annihilation processes can impact thermal evolution and cooling rates, potentially altering the emission signatures observed from compact stars \cite{10.1093/mnrasl/slv049, 10.1093/mnras/stac1013, 10.1093/mnras/stae337, PhysRevD.103.123022}. Understanding the interplay between nuclear matter and DM within NSs is crucial, as it directly affects their oscillatory behavior and stability. Since oscillations serve as a key probe for the internal structure of NSs, investigating how DM influences both radial and non-radial oscillation modes provides an important avenue for constraining the EOS and identifying potential signatures of DM.

Asteroseismology, the study of stellar oscillations, provides a powerful tool for probing the internal structure of NSs, much like helioseismology in the Sun and terrestrial seismology on Earth. In helioseismology, oscillations observed on the solar surface allow researchers to infer the Sun's internal composition and energy transport mechanisms \cite{RevModPhys.74.1073}. Asteroseismology of NSs follows the same principle-analyzing oscillation frequencies to extract information about their internal structure, EOS, and stability conditions \cite{PhysRevLett.77.4134, 10.1046/j.1365-8711.1998.01840.x, 10.1111/j.1365-2966.2006.11304.x}. The oscillations are broadly classified into radial and non-radial modes, distinguished by the nature of the perturbation. 

Radial oscillations pertain to the pulsations along the star's radius, preserving the star's spherical symmetry. These oscillations are intrinsically linked to the star’s stability and provide crucial insights into its structural integrity. Radial modes are governed by the linearized perturbation equations of hydrostatic equilibrium, forming an eigenvalue problem.
In this context, small radial perturbations to the pressure, density, and metric together with the radial displacement function, $\zeta_{r}$, lead to a Sturm-Liouville type second-order differential equation  \cite{Misner:1973prb, Kokkotas_2001}:
\begin{equation}
    \frac{d}{dr}\left({\cal P}(r)\frac{d\zeta_{r}}{dr}\right) +  \left[{\cal Q}(r)+\omega^{2}{\cal W}(r)\right]\zeta_{r} = 0,
\end{equation}
where ${\cal P}(r)$, ${\cal Q}(r)$, and ${\cal W}(r)$ are functions determined by the unperturbed stellar structure, and $\omega$ is the eigenfrequency of the radial mode. Solving this eigenvalue problem yields the characteristic frequencies of oscillation. A configuration is deemed dynamically stable if all eigenfrequencies satisfy $\omega^{2}>0$, while the presence of a negative $\omega^{2}$ implies the existence of an unstable mode that grows exponentially with time, signaling the onset of gravitational collapse. An alternative stability criterion arises from the turning-point method, which links stability to the dependence of stellar mass $M$ on the central energy density $\varepsilon_{c}$. Stability is maintained as long as $dM/d\varepsilon_{c} > 0$, while the point $dM/d\varepsilon_{c} = 0$ marks the onset of instability. This criterion, though simpler to evaluate, has been shown analytically to coincide with the vanishing of the fundamental mode frequency—signaling the onset of dynamical instability—as demonstrated in the general proof by Friedman, Ipser, and Sorkin (1988)~\cite{1988ApJ...325..722F} and supported by numerical results in \cite{1965gtgc.book.....H, 1996JApA...17..199F, Hadzic2021-vk}.
Beyond this point, i.e., for the stellar model whose central density is larger than that for the maximum stellar model, radial perturbations fail to restore equilibrium, and the star becomes dynamically unstable. The presence of DM alters the background equilibrium structure, which may shift the critical central density where instability sets in, either stabilizing or destabilizing the NS depending on the properties of the DM component and its coupling to nuclear matter. 

Non-radial oscillations, in contrast to radial oscillations, involve perturbations that include angular dependence and thus distort the spherical symmetry of the star. These modes are characterized by spherical harmonic indices and can be separated into two classes, depending on their parities, i.e., axial (odd-parity) and polar (even-parity) modes. Among these, polar modes are particularly important in the context of gravitational wave emission, as they involve fluid motions associated with density and pressure perturbations, leading to time-varying mass quadrupole moments and thus efficient coupling to spacetime curvature. Further, the polar modes
can be classified into several families depending on the nature of the restoring force. The most prominent of these are the fundamental ($f$-), pressure ($p$-), and gravity ($g$-) modes. The $f$-mode represents the lowest-order non-radial oscillation and is governed predominantly by the average density of the star, making it sensitive to both the EOS and global stellar structure. In contrast, the $p$-modes are restored by pressure gradients and typically manifest at higher frequencies, whereas the $g$-modes are driven by buoyancy due to the existence of stratified regions where the composition or entropy gradients are significant~\cite{1983ApJS...53...73L, 10.1093/mnras/227.2.265, 1990MNRAS.245..508M, PhysRevD.65.024010}. The study of non-radial modes, especially for the polar modes, has gained substantial relevance with the advent of gravitational wave astronomy. These oscillations couple to spacetime perturbations and can efficiently radiate gravitational waves. In particular, the $f$-modes are expected to be dominant in gravitational wave (GW) signals emitted by isolated NSs or post-merger remnants in binary star mergers. The detection of such signals would provide direct constraints on the internal structure and EOS of compact stars~\cite{Andersson_1998, 2002A&A...395..201Y, Heyl_2004, PhysRevD.69.124028}. Numerical calculations have shown that the frequencies of the $f$-modes lie in the kilohertz range and mass-scaled frequencies are well characterized by stellar compactness, making them promising targets for next-generation detectors~\cite{PhysRevD.86.063001, 2021PhRvD.103l3015S, GuhaRoy_2024}. In addition to their observational significance, non-radial oscillations also encode information about dynamical instabilities and energy dissipation mechanisms in NSs. For instance, the $r$-mode excited in a rotating star due to the Coriolis force shows instabilities in rapidly rotating stars, which can lead to long-duration GW emission, where their suppression or excitation is strongly influenced by the microphysics of dense matter and potential DM interactions \cite{1999A&A...341..110K, Yoshida:1999zp, PhysRevD.85.024007, Aasi_2015, Fesik_2020}. As such, understanding the full spectrum of non-radial modes, particularly in DM admixed NSs, is crucial for interpreting both continuous and transient GW signals.

In addition to the insights offered by stellar oscillation studies, another powerful diagnostic in NS physics arises from so-called universal relations—approximate, EOS-insensitive correlations among certain dimensionless macroscopic quantities of NSs. Prominent examples include the I-Love-Q relations, which connect the dimensionless moment of inertia ($I$), tidal deformability ($\Lambda$), and quadrupole moment ($Q$), and hold with remarkable accuracy across a wide class of cold, slowly rotating NS models \cite{PhysRevD.88.023009, 10.1093/mnras/stt858, PhysRevLett.123.051102}. Despite the large uncertainties in the nuclear EOS at supranuclear densities, these relations exhibit only weak dependence on microphysical details, making them highly valuable for astrophysical applications. These universal relations have been extended to binary systems as well, such as the binary Love relations linking the tidal deformabilities of the two stars in a merger event~\cite{PhysRevD.77.021502, Yagi_2016}. These relations have proven instrumental in extracting constraints from multimessenger observations like GW170817, where precise measurements of tidal deformability were used to infer stellar radii and test the validity of nuclear EOS models~\cite{PhysRevD.99.043010}. While these relations are largely EOS-independent for ordinary matter, the presence of exotic components like DM may lead to systematic deviations. Several recent works have investigated the extent to which DM—especially when modeled as a separate fluid or as interacting particles within a single-fluid framework—modifies the universal behavior of NSs \cite{PhysRevD.110.103033, PhysRevD.104.123006}. Studying the robustness and possible breakdown of these relations in DM admixed NSs provides an important diagnostic for the detectability of DM effects and could serve as an indirect probe for the presence and distribution of DM in compact stars.

In this work, we investigate the impact of DM on the oscillation modes and universal relations of NSs, using a single-fluid formalism in which DM interacts with nuclear matter via an effective EOS. We construct the background equilibrium configuration by solving the general relativistic stellar structure equations and examine how the presence of DM modifies the mass-radius relation. We then analyze radial oscillations to probe the stability of these configurations and also compute the fundamental frequencies of non-radial oscillations relevant for GW emission. Finally, we test the robustness of several proposed universal relations in the presence of DM. This paper is organized as follows. In Sec.~\ref{sec:2}, we describe the EOS models used for nuclear and DM and present the background equilibrium structure. In Sec.~\ref{sec:3}, we outline the formalism for radial oscillations and present the resulting stability analysis. In Sec.~\ref{sec:4}, we extend the study to non-radial $f$-mode oscillations and examine their DM dependence. In Sec.~\ref{sec:5}, we explore the validity and breakdown of universal relations in DM admixed NSs. Finally, we summarize our findings in Sec.~\ref{sec:6}.

%%%%%%%%%%%%%%%%%%%%%%%%%%%%%
\section{Equation of state and background for perturbations}
\label{sec:2}
%%%%%%%%%%%%%%%%%%%%%%%%%%%%%
To model the interior structure of DM admixed NSs, we adopt a single-fluid formalism in which DM interacts with nuclear matter through an effective EOS. The total energy density and pressure in this approach are functions of both nuclear and DM contributions, treated self-consistently within the relativistic framework.
%%%%%%%%%%%%%%%%%%%%%%%%%%%%%
% Figure 1
%%%%%%%%%%%%%%%%%%%%%%%%%%%%%
\begin{figure*}[tbp]
    \centering
    \includegraphics[width=\textwidth]{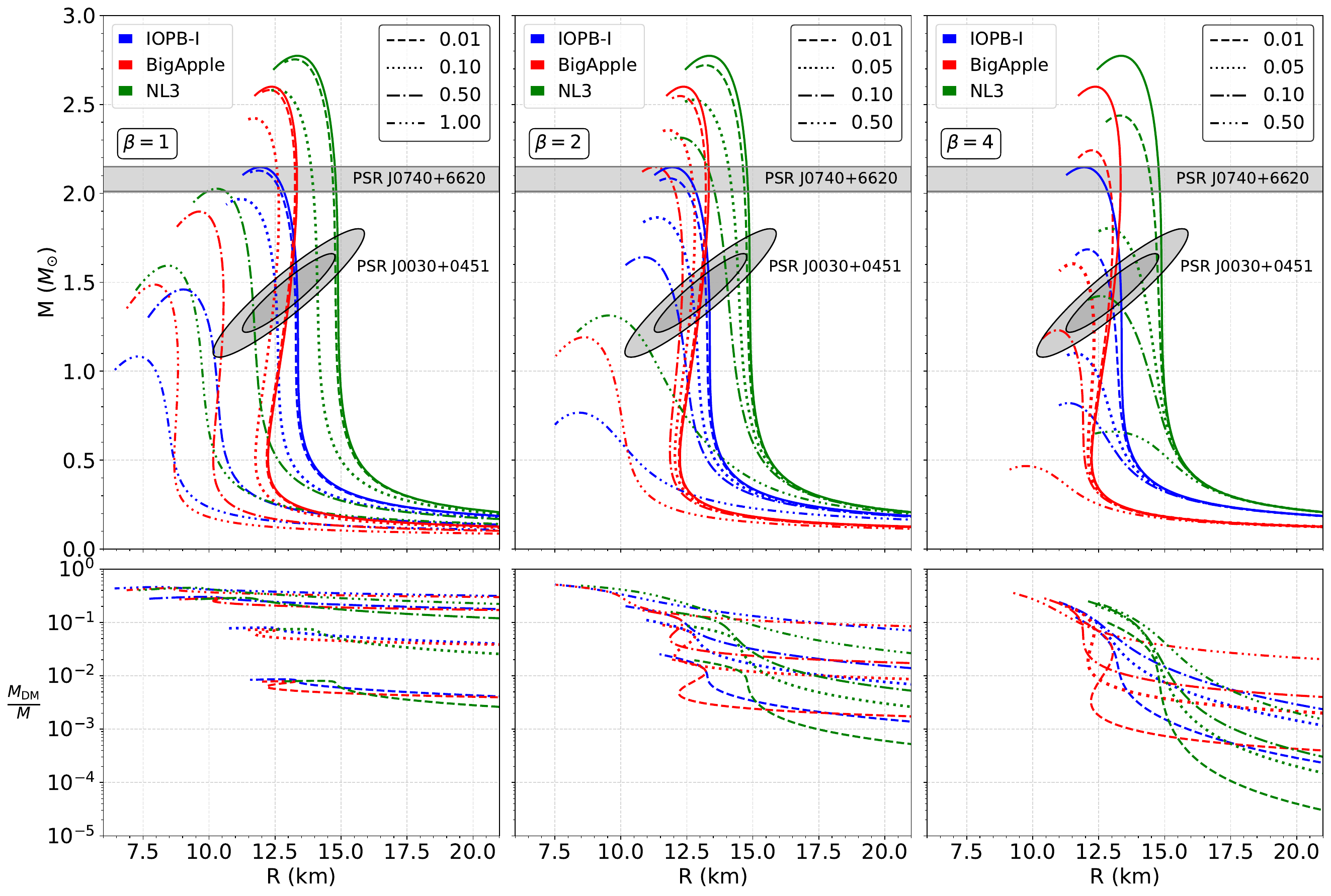}
    \caption{The top panels display the mass-radius relations for NSs constructed using the NL3, BigApple, and IOPB-I RMF parameter sets, incorporating DM with a variable density profile for steepness values $\beta=1$ (left), $\beta=2$ (middle), and $\beta=4$ (right). Each curve corresponds to a fixed value of $\alpha M_{\chi}$ from 0.01 up to 1.00, highlighting the dominant role of this product in controlling the DM gravitational influence. In each panel, the solid line corresponds to the result without DM. Observational constraints from NICER analysis for PSR J0030+0451 \cite{Riley_2019, Vinciguerra_2024} and PSR J0740+6620 \cite{Fonseca_2021, Dittmann_2024} are shown for reference. The bottom panels display the corresponding DM mass fraction, $M_{\rm{DM}}/M$, as a function of radius, providing a complementary view of the DM contribution across the stellar interior for each EOS and corresponding combinations of ($\alpha M_{\chi}$, $\beta$) parameter space.} 
    \label{fig:figure1}
\end{figure*}
%%%%%%%%%%%%%%%%%%%%%%%%%%%%%

For the nuclear matter sector, we employ the relativistic mean-field (RMF) theory, a well-established framework in nuclear astrophysics that describes dense matter through meson-exchange interactions between baryons. Although the RMF Lagrangian is not reproduced here, it consists of nucleons interacting via scalar ($\sigma$), vector ($\omega$), and isovector ($\rho$) mesons, and has been extensively detailed in prior literature~\cite{WALECKA1974491, 1996csnp.book.....G, 1977NuPhA.292..413B, FURNSTAHL1997441, PhysRevC.85.024302, PhysRevC.55.540, KUBIS1997191, PhysRevC.62.015802, Kumar2020}. To account for the uncertainties in the high-density behavior of nuclear matter, we adopt three distinct parameter sets within the RMF framework:
\begin{itemize}
    \item NL3 \cite{PhysRevC.55.540} — a stiff EOS characterized by high incompressibility and a large symmetry energy slope, leading to large NS radii and high maximum masses.
    \item BigApple \cite{PhysRevC.102.065805} — a moderately stiff parameter set and predicts maximum masses up to $\sim 2.6 M_{\odot}$, potentially accommodating the secondary component of GW190814 \cite{Abbott_2020}, whose nature—whether a massive NS or a light black hole—remains under debate. Notably, BigApple satisfies nuclear saturation properties, making it somewhat unconventional, as relativistic EOSs that yield such high masses typically fail to reproduce empirical constraints from finite nuclei and nuclear matter experiments.
    \item IOPB-I \cite{PhysRevC.97.045806} — represents a softer EOS, calibrated to recent nuclear physics data and astrophysical observations, providing more compact stellar configurations consistent with constraints from NICER measurements of PSR J0740+6620 \cite{Fonseca_2021, Dittmann_2024} and PSR J0030+0451 \cite{Riley_2019, Vinciguerra_2024}.
\end{itemize}
These parameter sets span a broad range of stiffness, allowing us to examine the sensitivity of oscillation modes and universal relations to the nuclear EOS. 

For the DM sector, we adopt a Higgs-mediated fermionic DM model. The corresponding Lagrangian is given by \cite{PhysRevD.96.083004}
%%%%%%%%%%%%%%%%%%%%%%%%%%%%%
\begin{align}
    \mathcal{L}_{\rm DM} =& \,\bar\chi\left[i\gamma^\mu \partial_\mu - M_\chi + y h \right]\chi 
    \nonumber \\
    & + \frac{1}{2}\partial_\mu h \partial^\mu h - \frac{1}{2} M_h^2 h^2 + \frac{f\,M_{n}}{v} \bar \psi h \psi,   
    \label{eq:L_DM}
\end{align}
%%%%%%%%%%%%%%%%%%%%%%%%%%%%%
where $\chi$ is the DM fermion with mass $M_{\chi}$, and $h$ is the Higgs field with mass $M_{h}$. The final term represents the coupling between the Higgs and standard model nucleons $\psi$ with the nucleon mass $M_{n}$ and the Higgs vacuum expectation value, $v = 246$ GeV. The coupling parameters $y$ and $f$, which govern the Yukawa coupling between DM and Higgs and the strength of Higgs-nucleon interactions, respectively, are adopted following the values used in Refs.~\cite{PhysRevD.96.083004, PhysRevD.110.063001}. This interaction enables DM to influence the nuclear EOS indirectly through Higgs exchange.

This Higgs-portal DM model has been previously explored in the context of NSs in several studies \cite{PhysRevD.99.043016, PhysRevD.104.063028, 10.1093/mnras/stab2387, Routaray_2023, ABAC2023101185, Routaray_2023, PhysRevD.107.103039, PhysRevD.109.083021, PhysRevD.110.063001}, where a simplifying assumption of constant DM Fermi momentum—or equivalently, a uniform DM density throughout the star—was commonly employed. While analytically convenient, this approximation is physically inconsistent, as such an assumption neglects the effect of the NS's intense gravitational field, which should naturally lead to a more centrally concentrated DM distribution. A more realistic treatment considers the gravitational trapping of DM, resulting in a density profile that peaks at the core and diminishes toward the outer layers. To capture this behavior, we adopt a variable DM number density profile introduced in our previous work \cite{PhysRevD.111.043016}, parameterized as
%%%%%%%%%%%%%%%%%%%%%%%%%%%%%
\begin{align}
   \frac{n_{\rm DM}}{n_{0}} = \alpha \Bigg(\frac{n_{\rm B}-n_t}{n_{0}}\Bigg)^{\beta},
   \label{eq:ndensity}
\end{align}
%%%%%%%%%%%%%%%%%%%%%%%%%%%%%
where $n_{\rm{DM}}$ is the local DM number density, $n_{0}$ is the nuclear saturation density, and $n_{\rm B}$ is the baryon number density. The parameter $\alpha$ controls the overall DM fraction (scaling), while $\beta$ determines the steepness of the density profile. The threshold density $n_{\rm t}$ defines the minimum baryon density required for the presence of DM, and in this work, it is identified with the core-crust transition density. This choice ensures that DM is confined to the high-density core and does not populate the crustal region, in line with the expectation that gravitational capture of weakly interacting particles is effective only in the dense interior. For the crust region, we employ the SLy4a EOS~\cite{refId0, CHABANAT1998231}, ensuring a consistent treatment of the low-density regime where baryonic matter dominates and DM is negligible. We explicitly set $n_{\rm{DM}}$ for the region where the baryon number density falls below the threshold $n_{t}$, ensuring that dark matter is present only in the dense interior of the star.

The equilibrium structure is obtained by solving the Einstein field equations under the assumption of a static, spherically symmetric spacetime. The line element is expressed as
%%%%%%%%%%%%%%%%%%%%%%%%%%%%%
\begin{equation}
    ds^{2} = -e^{2\Phi(r)} dt^{2} + e^{2\lambda(r)} dr^{2} + r^{2} d\theta^{2} + r^{2} \sin^{2}\theta\, d\phi^{2},
    \label{eq:linement}
\end{equation}
%%%%%%%%%%%%%%%%%%%%%%%%%%%%%
where $\Phi(r)$ and $\lambda(r)$ are metric functions determined by the stellar mass and pressure distributions. The metric function $e^{2\Lambda}$ satisfies the relation $e^{-2\lambda} = 1-2m/r$, where $m(r)$ is the enclosed gravitational mass. The system of equations governing the static equilibrium configuration is given by:
%%%%%%%%%%%%%%%%%%%%%%%%%%%%%
\begin{align}
    \frac{d\Phi}{dr} &= \ \frac{m+4\pi r^{3}p}{r(r-2m)},  \\
    \frac{dm}{dr} &= \ 4\pi r^{2} \varepsilon,  \\
    \frac{dp}{dr} &= \ -\left(\varepsilon + p\right) \frac{d\Phi}{dr},
\end{align}
%%%%%%%%%%%%%%%%%%%%%%%%%%%%%
where $\varepsilon(r)$ and $p(r)$ are the total energy density and pressure at radius $r$, incorporating both nuclear and DM contributions. These equations resemble the standard Tolman-Oppenheimer-Volkoff (TOV) equations, but are expressed in a form that emphasizes the gravitational potential $\Phi(r)$, rather than relying solely on the pressure gradient and mass function. In particular, the adopted set of equations highlights the role of $\Phi(r)$ as an explicit dynamical variable, which is especially useful when solving the perturbation equations for oscillation modes. We refer to this system as the background for solving the subsequent perturbation equations. The integration of these equations from the center ($r=0$) to the surface ($p(R)=0$) yields the global properties of the star, such as total mass $M=m(R)$ and radius $R$. These background profiles form the foundation for analyzing both radial and non-radial oscillations in the presence of DM.

Using these equations for a background model, we compute the mass-radius ($M-R$) profiles for NSs constructed with the three RMF parameter sets—NL3, BigApple, and IOPB-I—for fixed values of the DM steepness parameter $\beta = 1, 2$ and 4, as shown in Fig.~\ref{fig:figure1}. For each $\beta$, we explore a range of $\alpha M_{\chi}$ values, where $\alpha$ is the scaling factor in the DM number density profile, and $M_{\chi}$ is the DM particle mass. This choice is motivated by our earlier findings \cite{PhysRevD.111.043016}, which show that the product $\alpha M_{\chi}$ primarily controls the total DM energy density and hence its gravitational effect on the star. At fixed $\beta$, we observe that models with the same $\alpha M_{\chi}$ values produce identical mass-radius curves, independent of the individual values of $\alpha$ or $M_{\chi}$, emphasizing the relevance of $\alpha M_{\chi}$ as the effective controlling parameter.

In addition to the mass-radius relations, the lower panels of Fig.~\ref{fig:figure1} present the corresponding DM mass fraction profiles, expressed as $M_{\rm{DM}}/M$, as a function of stellar radius. Here, $M_{\rm{DM}}$ denotes the gravitational mass composed of the DM component, and $M$ is the total gravitational mass of the configuration. These plots provide complementary insight into the spatial distribution and relative contribution of DM within the star, particularly highlighting how the DM fraction evolves with radius for different EOSs and parameter combinations. Notably, while the gravitational imprint of DM on the global mass-radius relation is largely governed by $\alpha M_{\chi}$, the DM mass fraction profiles reveal the localized impact of DM across the stellar interior.

Increasing $\alpha M_{\chi}$ results in a systematic decrease in radius for a given mass, reflecting enhanced gravitational confinement from the DM component. The effect is more pronounced for larger $\beta$, as the DM becomes increasingly localized near the stellar center. This highlights the sensitivity of DM concentration (or distribution) in determining the structural properties of the star. In addition to shrinking the radius, the intense concentration of DM in the core also leads to a reduction in the maximum mass. Notably, for sufficiently large $\alpha M_{\chi}$, the maximum mass may fall below the observational lower bound set by massive pulsars such as PSR J0740+6620. A detailed discussion about the gravitational confinement of DM, its impact on stellar structure, and the examination of the allowed ranges for $\alpha M_{\chi}$ and $\beta$ consistent with observational data is presented in our earlier work \cite{PhysRevD.111.043016}.
%%%%%%%%%%%%%%%%%%%%%%%%%%%%%
%%%%%%%%%%%%%%%%%%%%%%%%%%%%%
%%%%%%%%%%%%%%%%%%%%%%%%%%%%%
%%%%%%%%%%%%%%%%%%%%%%%%%%%%%
%%%%%%%%%%%%%%%%%%%%%%%%%%%%%
%%%%%%%%%%%%%%%%%%%%%%%%%%%%%
%%%%%%%%%%%%%%%%%%%%%%%%%%%%%
%%%%%%%%%%%%%%%%%%%%%%%%%%%%%
%%%%%%%%%%%%%%%%%%%%%%%%%%%%%
%%%%%%%%%%%%%%%%%%%%%%%%%%%%%
%%%%%%%%%%%%%%%%%%%%%%%%%%%%%
\section{Radial Oscillations and Stability Analysis}
\label{sec:3}
%%%%%%%%%%%%%%%%%%%%%%%%%%%%%
The study of radial oscillations in relativistic stars provides a crucial window into their dynamical stability and internal composition. The earliest systematic treatment of relativistic radial pulsations was developed by Chandrasekhar in seminal work \cite{1964ApJ...140..417C}, where he derived a second-order Sturm-Liouville eigenvalue equation for the radial displacement variable under linear perturbations of a static, spherically symmetric background. His formulation allowed for determining stability by analyzing the sign of the eigenfrequencies: configurations with only real, positive squared frequencies are dynamically stable, while a negative eigenvalue indicates an unstable mode.

Subsequently, several researchers reformulated the radial perturbation problem to enable more robust numerical implementations. Notably, Chanmugam~\cite{1977ApJ...217..799C} introduced a set of two first-order coupled differential equations involving the Lagrangian radial displacement $\xi(r)$ and a pressure-related perturbation function $\eta(r)$, which has since become a standard approach in numerical studies. Although the radial oscillation problem can be formulated as a second-order Sturm-Liouville equation for the displacement function $\xi(r)$, it is customary to rewrite it as a first-order system involving $\xi(r)$ and the Lagrangian pressure perturbation. This approach, introduced by Chanmugam, simplifies the imposition of boundary conditions and improves numerical tractability, particularly near the stellar surface where second-order formulations may suffer from numerical instabilities. Further refinements and numerical strategies have been explored in the literature~\cite{1997A&A...325..217G, Kokkotas_2001}.

In this work, we adopt the first-order formalism presented in~\cite{10.1093/mnras/stab2301}, where the radial oscillation equations are expressed as:
%%%%%%%%%%%%%%%%%%%%%%%%%%%%%
\begin{eqnarray}
    \frac{d\xi}{dr} &=& - \left[\frac{3}{r}+\frac{p'}{p+\varepsilon}\right]\xi - \frac{\eta}{r\Gamma} \\
    \frac{d\eta}{dr} &=& \left[r\left(p+\varepsilon\right)e^{2\lambda}\left(\frac{\omega^{2}}{p}e^{-2\Phi}-8\pi\right)-\frac{4p'}{p} + \frac{r(p')^{2}}{p\left(p+\varepsilon\right)}\right] \xi \nonumber \\
    &&  - \left[\frac{\varepsilon p'}{p\left(p+\varepsilon\right)} + 4\pi r\left(p + \varepsilon\right)e^{2\lambda}\right]\eta
    \label{eq:radperteqs}
\end{eqnarray}
%%%%%%%%%%%%%%%%%%%%%%%%%%%%%
where $p' = dp/dr$ and $\Gamma$ is the adiabatic index defined as
\begin{equation}
  \Gamma \equiv \frac{\varepsilon + p}{p} \left(\frac{\partial p}{\partial \varepsilon}\right)_{s},
\end{equation}
with the subscript $s$ denoting that the derivative is evaluated at constant entropy, consistent with the assumption of adiabatic (isentropic) perturbations. The functions $\xi(r)$ and $\eta(r)$ respectively describe the radial displacement of fluid elements and the associated Lagrangian perturbation in pressure. The oscillation frequency $\omega$ enters as an eigenvalue in the system and determines the stability of the configuration. 
%%%%%%%%%%%%%%%%%%%%%%%%%%%%%
% Figure 2
%%%%%%%%%%%%%%%%%%%%%%%%%%%%%
\begin{figure}[htbp]
    \centering
    \includegraphics[width=\columnwidth]{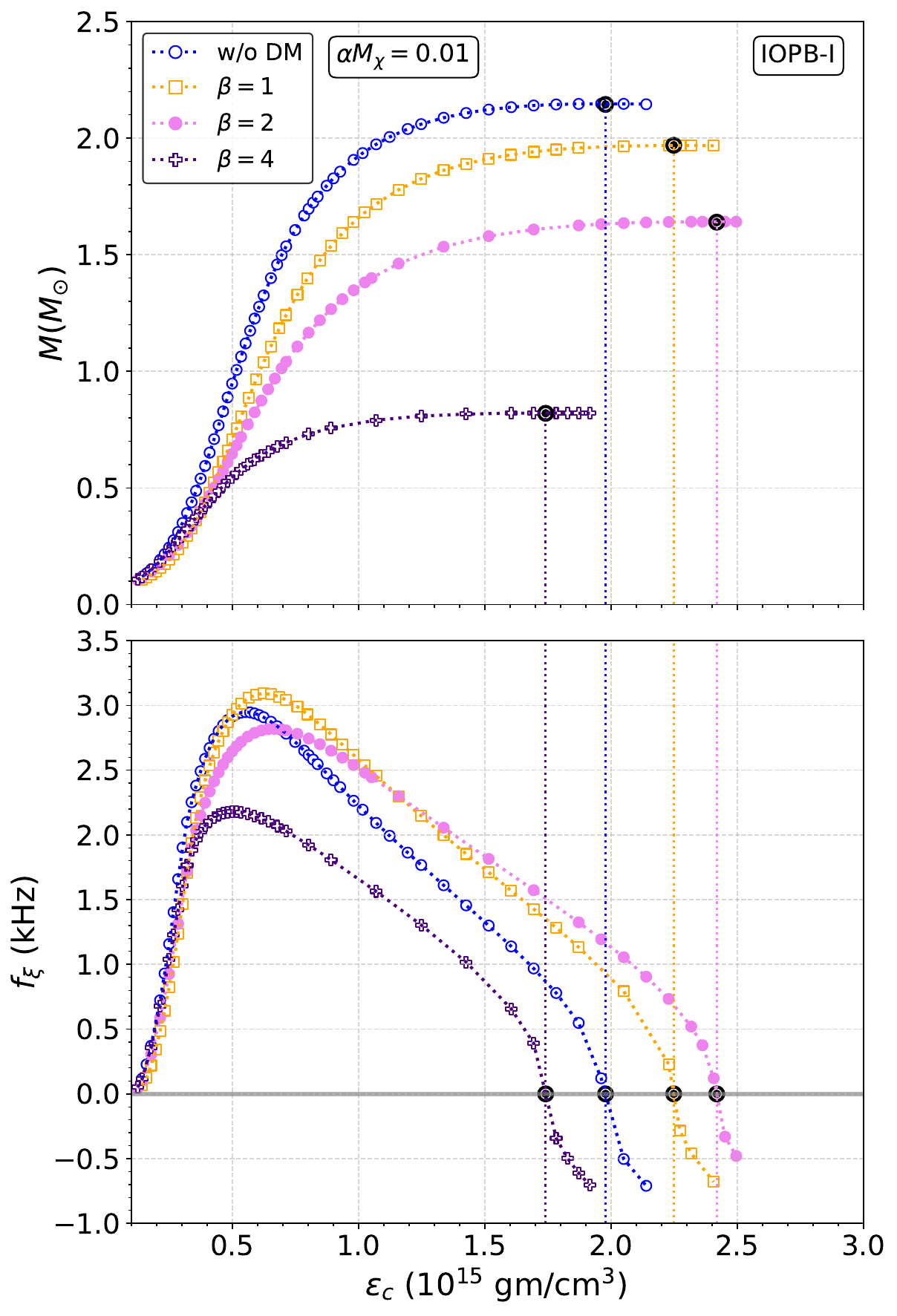}
    \caption{Frequency of the fundamental radial oscillation, $f_\xi$ (in kHz), as a function of central energy density $\varepsilon_c$ (in $10^{15}$ g/cm$^3$) for the IOPB-I EOS. The curve labeled “w/o DM” corresponds to the NS configuration with pure baryonic component, while the others show results for DM-admixed configurations with fixed $\alpha M_\chi = 0.01$ and steepness parameter $\beta = 1$, 2, and 4. In each case, the point where $f_\xi \to 0$ marks the onset of dynamical instability and coincides with the configuration of maximum mass. The figure illustrates how DM concentration impacts both the frequency spectrum and the critical central density for stability.
    }
    \label{fig:figure2}
\end{figure}
%%%%%%%%%%%%%%%%%%%%%%%%%%%%%
%%%%%%%%%%%%%%%%%%%%%%%%%%%%%
% Figure 3
%%%%%%%%%%%%%%%%%%%%%%%%%%%%%
\begin{figure*}[tbp]
    \centering
    \includegraphics[width=\textwidth]{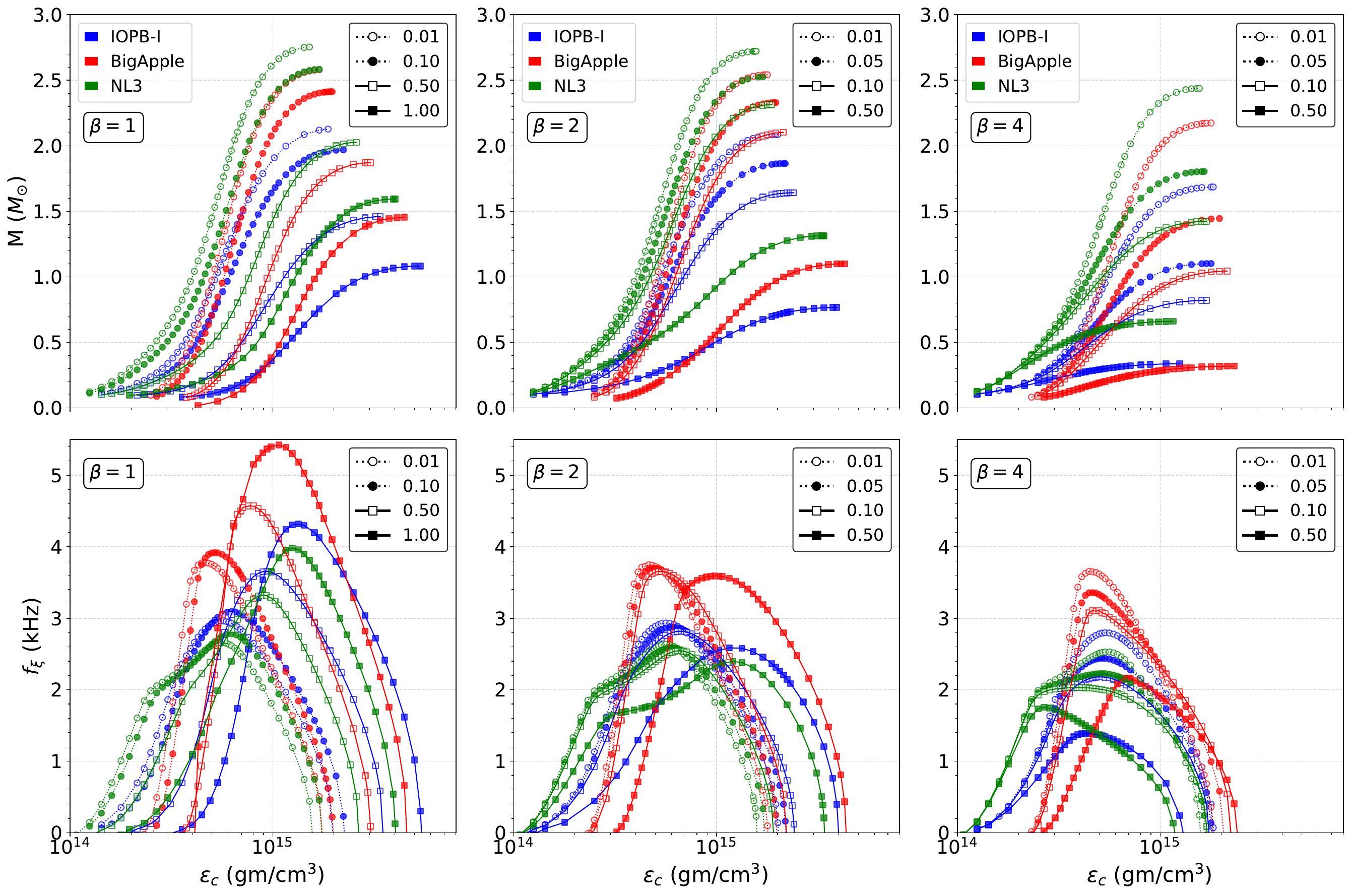}
    \caption{Stellar mass (top row) and the frequency of the fundamental radial oscillation  $f_{\xi}$ (bottom row) as functions of central energy density $\varepsilon_{c}$ for three RMF nuclear matter EOS models: IOPB-I (blue), BigApple (red), and NL3 (green). Each column corresponds to a fixed steepness parameter: $\beta=1$ (left), $\beta=2$ (middle), and $\beta=4$ (right). Within each panel, different marker styles represent different values of the DM scaling parameter $\alpha M_{\chi}$, as indicated in the legend. The curves are truncated at the onset of instability, and only configurations with positive squared oscillation frequencies ($\omega^{2} > 0$) are displayed for each case.}
    \label{fig:figure3}
\end{figure*}
%%%%%%%%%%%%%%%%%%%%%%%%%%%%%
To obtain physically meaningful solutions, appropriate boundary and regularity conditions must be imposed:
%%%%%%%%%%%%%%%%%%%%%%%%%%%%%
\begin{itemize}
    \item Center ($r=0$) : Regularity demands that the displacement function behaves smoothly as $r \rightarrow 0$, leading to the condition: \begin{equation} 3\Gamma\xi + \eta = 0. \end{equation} We normalize the solution by choosing $\xi(0) = 1$, as the overall amplitude is arbitrary in linear perturbation theory.
    \item Surface ($r = R$) : At the stellar surface, the Lagrangian pressure perturbation must vanish, i.e., $\Delta p = 0$, which translates to the boundary condition given by \begin{align} \eta = -\left[\left(\frac{\omega^{2}R^{3}}{M}+\frac{M}{R}\right)\left(1-\frac{2M}{R}\right)^{-1}+4\right]\xi\ . \end{align}
\end{itemize}
%%%%%%%%%%%%%%%%%%%%%%%%%%%%%
The eigenfrequencies $\omega^2$ are then determined by solving this boundary value problem. Positive $\omega^2$ indicates stable oscillatory modes, while negative $\omega^2$ corresponds to the dynamically unstable sellar models. Finally, for reference and comparison, we compute the corresponding characteristic oscillation frequency $f_\xi$ in physical units as:
%%%%%%%%%%%%%%%%%%%%%%%%%%%%%
\begin{equation}
    f_{\xi} = \rm{sign}\left(\omega^{2}\right)\ \frac{\sqrt{|\omega^{2}|}}{2\pi}.
\end{equation}
%%%%%%%%%%%%%%%%%%%%%%%%%%%%%
We note that the use of a single-fluid formalism in this work is justified by the assumption that DM interacts with nuclear matter via an effective Higgs-mediated coupling, leading to a unified equation of state. The local dark matter number density is directly tied to the baryon density via Eq. \eqref{eq:ndensity}, ensuring that the dark matter component is spatially and dynamically coupled to the nuclear sector. While we do not compute the thermalization or interaction timescales explicitly, we assume that the two components respond coherently to small fluid perturbations within this effective framework. As such, the single-fluid approximation is adopted as a modeling choice to explore the regime where dark matter remains co-evolving with the baryonic matter during stellar oscillations. In Fig.~\ref{fig:figure2}, we show the frequency of the fundamental radial oscillation, $f_\xi$, as a function of central energy density, $\varepsilon_c$, for stellar models constructed with the IOPB-I EOS. The results are shown for the star composed of pure nuclear matter (denoted "w/o DM") and for DM-admixed stars with fixed $\alpha M_{\chi} = 0.01$ and steepness parameters $\beta = 1$, 2, and 4. As expected, the characteristic fundamental frequency $f_{\xi}$ remains positive over the stable branch of stellar configurations, indicating dynamical stability. As the central density increases, the fundamental frequency $f_\xi$ decreases and vanishes at a critical point, marking the onset of instability. This transition corresponds to the maximum mass configuration in the mass-central density relation even for the DM admixed NSs and reflects the fundamental mode frequency becoming zero ($\omega^2 \to 0$), in accordance with the turning-point criterion discussed earlier.

While the presence of DM modifies the overall stellar structure, it does not shift the criterion for stability: the onset of instability always coincides with the maximum mass for a given $\alpha M_\chi$ and $\beta$. The maximum mass decreases monotonically with increasing $\beta$, reflecting stronger gravitational confinement from more localized DM distributions. In contrast, the critical central density shows non-monotonic behavior: increasing $\beta$ from 1 to 2 pushes the instability point to higher $\varepsilon_c$, whereas for $\beta = 4$, the sharply peaked DM profile near the core reduces its effective pressure support, leading to a lower critical $\varepsilon_c$. This trend reflects a subtle competition between gravitational confinement and spatial distribution of DM in shaping stellar stability.

Figure~\ref{fig:figure3} presents the variation of stellar mass (top row) and the frequencies of the fundamental radial oscillations  $f_{\xi}$ (bottom row) as functions of central energy density $\varepsilon_{c}$, for three nuclear matter EOS models: IOPB-I, BigApple, and NL3. The three columns correspond to different values of the DM density steepness parameter: $\beta =1 , 2$, and 4 from left to right. Within each panel, different values of the DM scaling parameter $\alpha M_{\chi}$ are represented using distinct marker styles (filled/open circle and square markers), as indicated in the legend. For each configuration, the curves are terminated at the stellar model with maximum mass in the top panels and at the onset of instability in the bottom panels, defined by the point where $\omega^{2} = 0$, i.e., only the dynamically stable configurations ($\omega^{2} > 0$) are shown.

For each EOS and fixed $\beta$, increasing $\alpha M_{\chi}$ leads to a systematic reduction in the maximum mass, as shown in the top row. This effect arises from the added gravitational contribution of the DM component, which modifies the internal pressure balance. Among the three EOS models, NL3 consistently predicts the highest maximum mass due to its inherent high stiffness, followed by BigApple and IOPB-I. However, the critical central energy density corresponding to the stellar model with the maximum mass varies depending on both the EOS and the DM parameters.

The bottom row shows the corresponding frequency of the fundamental radial oscillation $f_{\xi}$ as a function of $\varepsilon_c$. The frequency does not decrease monotonically with increasing $\varepsilon_c$; instead, it exhibits a nontrivial dependence: starting from a small value, $f_{\xi}$ initially increases with central density, reaches a peak, and then gradually decreases to zero at the instability point. This behavior is consistent with the known sequence of $f_{\xi}$ in the stable branch and highlights the sensitivity of $f_{\xi}$ to the internal structure of the star.

A closer examination of the peak values of $f_{\xi}$ reveals a subtle but consistent dependence on the interplay between the DM scaling parameter $\alpha M_{\chi}$ and the steepness parameter $\beta$, as governed by the DM density profile in Eq.~\eqref{eq:ndensity}. This profile implies that the local DM number density is proportional to a power-law function of the baryon number density, modulated by the parameters $\alpha$ and $\beta$. 

For a fixed EOS and $\beta = 1$, larger values of $\alpha M_{\chi}$ yield higher peak frequencies. At low $\beta$, the DM profile grows gently with baryon density, resulting in a relatively broad and smoothly distributed DM core. Increasing $\alpha$ enhances the overall DM energy content, leading to greater gravitational compression, steeper pressure gradients, and higher central compactness. These effects strengthen the restoring force acting on fluid elements and raise the peak value of $f_{\xi}$, despite the associated reduction in stellar mass with $\alpha M_{\chi}$.

As $\beta$ increases, the profile steepens, and the DM distribution becomes increasingly sensitive to the inner structure of the star. At $\beta = 2$, the dependence of peak frequency on $\alpha M_{\chi}$ weakens, and different configurations begin to exhibit comparable values of $f_{\xi}$. At $\beta = 4$, the trend reverses: smaller values of $\alpha M_{\chi}$ now correspond to higher peak frequencies. In this high-$\beta$ regime, the DM density becomes extremely steep, concentrating sharply near the core. For large $\alpha M_{\chi}$, although the total DM content is higher, it gets packed into an extremely narrow central region. As a result, its influence is limited to a small volume and does not effectively modify the pressure gradient across the star. This reduces its contribution to the restoring force responsible for radial oscillations. On the other hand, smaller $\alpha M_{\chi}$ values lead to a less concentrated DM profile that spreads more broadly through the core. This wider spatial support enhances pressure gradients over a larger region, providing stronger restoring forces and thus yielding higher peak frequencies.

Thus, the observed non-monotonic trend of $f_{\xi}$ with $\alpha M_{\chi}$ across increasing $\beta$ values is a direct manifestation of how gravitational confinement and spatial extent of DM influence the internal force balance. It reflects the complex interplay between total DM content and its spatial distribution in shaping the dynamical response of NSs to radial oscillations.

Another important observation is that the frequency of the radial oscillation  $f_{\xi}$ does not scale straightforwardly with the stellar mass or compactness. While NL3 consistently supports the most massive configurations due to its stiffness, it does not always yield the highest $f_{\xi}$. For low-mass stars—i.e., at lower central densities—NL3 parameter set typically exhibits higher oscillation frequencies than BigApple. However, as the central density increases, BigApple tends to overtake NL3 and produce larger $f_{\xi}$ values across much of the stable branch. This behavior indicates that $f_{\xi}$ is not solely governed by global properties such as mass, compactness, or star's average density, but is strongly influenced by the internal structure of the EOS, particularly the pressure gradient in the stellar core. While the ordering $f_{\xi}^{\rm BigApple} > f_{\xi}^{\rm IOPB-I} > f_{\xi}^{\rm NL3}$ does not persist throughout the entire stable sequence, it is clearly observed in the intermediate-$\varepsilon_{c}$ regions of the fundamental mode stable branch, reflecting the role of EOS microphysics in shaping the radial oscillation spectrum.
%%%%%%%%%%%%%%%%%%%%%%%%%%%%%
%%%%%%%%%%%%%%%%%%%%%%%%%%%%%
\section{Non-Radial Oscillations}
\label{sec:4}
%%%%%%%%%%%%%%%%%%%%%%%%%%%%%
%%%%%%%%%%%%%%%%%%%%%%%%%%%%%
% Figure 4
%%%%%%%%%%%%%%%%%%%%%%%%%%%%%
\begin{figure*}[htbp]
    \centering
    \includegraphics[width=\textwidth]{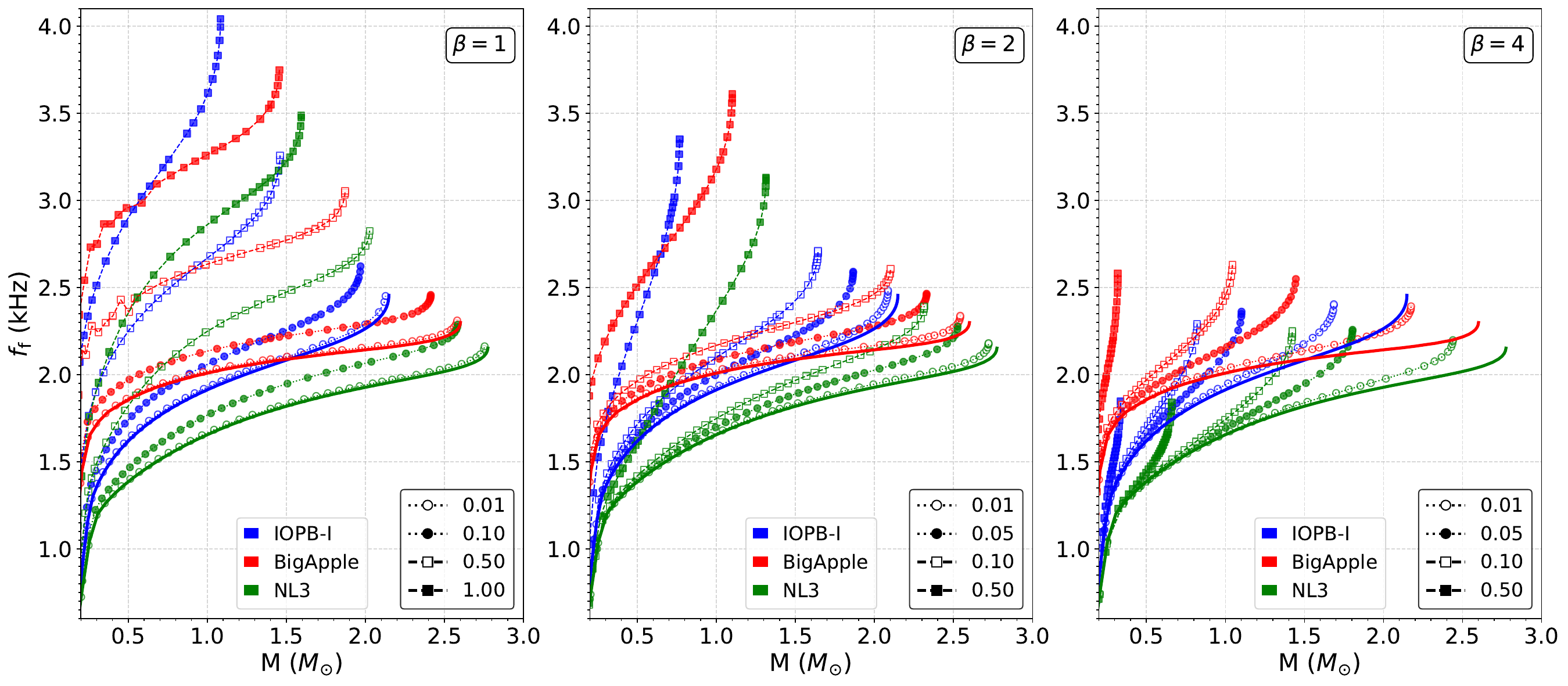}
    \caption{Frequency of the fundamental non-radial ($f$-mode) oscillation $f_{\rm f}$ (in kHz) as a function of stellar mass $M$ (in units of $M_{\odot}$) for NSs modeled with three different nuclear matter EOS: IOPB-I (blue), BigApple (red), and NL3 (green). Each panel corresponds to a fixed value of the DM density steepness parameter $\beta = 1, 2$, and 4 (from left to right). Solid lines represent the configurations without any DM component (i.e., pure baryonic stars), while the dashed curves correspond to DM-admixed stars with different values of the DM scaling parameter $\alpha M_{\chi}$, distinguished by marker styles as indicated in the legend. All sequences are truncated at the maximum mass configuration, which also marks the onset of dynamical instability, as discussed in the previous section.}
    \label{fig:figure4}
\end{figure*}
%%%%%%%%%%%%%%%%%%%%%%%%%%%%%
%%%%%%%%%%%%%%%%%%%%%%%%%%%%%
% Figure 5
%%%%%%%%%%%%%%%%%%%%%%%%%%%%%
\begin{figure*}[htbp]
    \centering
    \includegraphics[width=\textwidth]{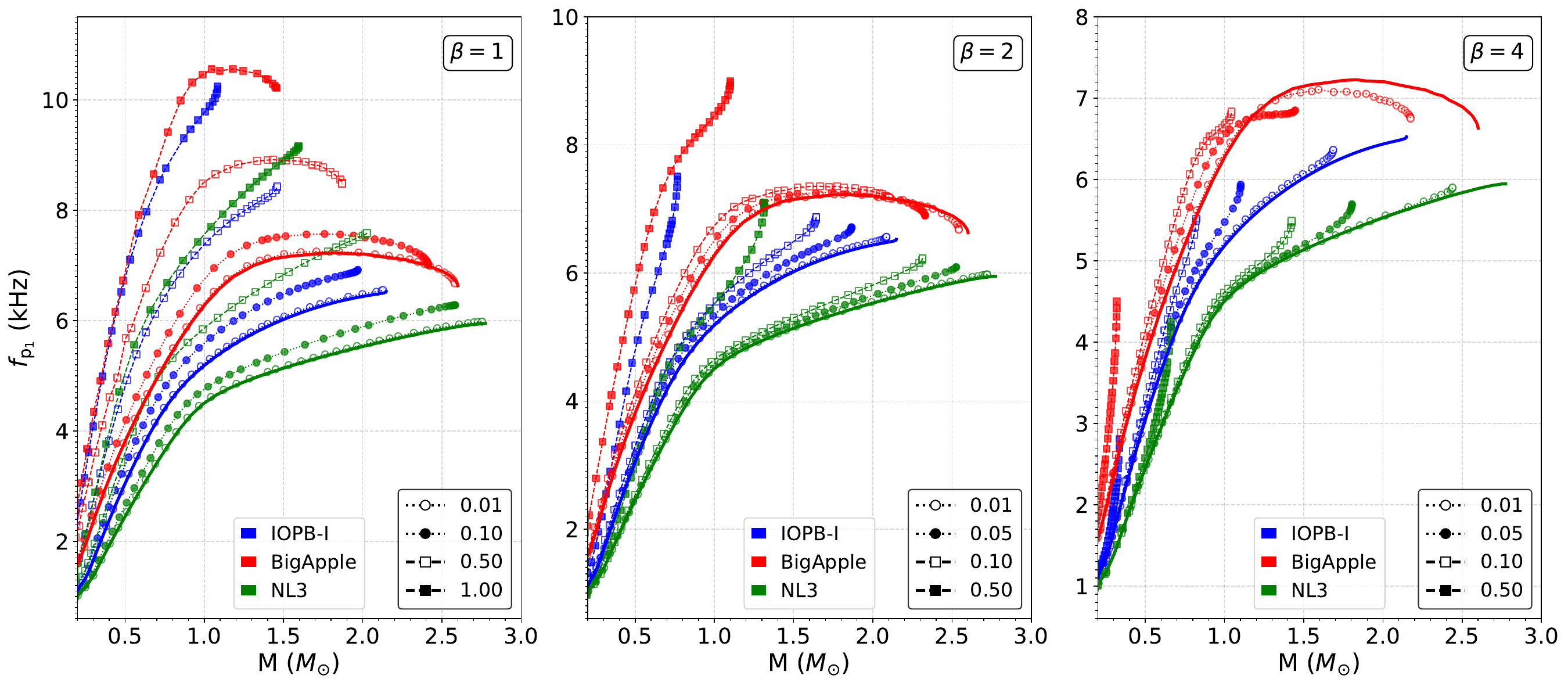}
    \caption{Same as Fig.~\ref{fig:figure4}, but for the oscillation frequency of the first pressure ($p_1$-) mode, $f_{p_1}$ (in kHz), plotted against stellar mass.}
    \label{fig:figure5}
\end{figure*}
%%%%%%%%%%%%%%%%%%%%%%%%%%%%%
Non-radial oscillations of NSs provide another important channel for probing their internal composition, offering a rich spectrum of modes that can couple to gravitational radiation. In full general relativity, the study of these oscillations involves solving the linearized Einstein field equations coupled with the conservation of energy-momentum for the fluid. The pioneering works by Thorne and Campolattaro \cite{1967ApJ...149..591T} and Lindblom and Detweiler \cite{1985ApJ...292...12D} laid the foundation for such analyses, establishing the classification of fluid modes (such as $f$-, $p$-, and $g$-modes) and the role of spacetime perturbations in gravitational wave emission.

However, solving the full system of perturbed Einstein-fluid equations is computationally demanding. A widely adopted simplification is the relativistic Cowling approximation~\cite{10.1093/mnras/101.8.367, 1983ApJ...268..837M}, in which perturbations to the spacetime metric are neglected, and only fluid variables are evolved. This approximation is particularly effective for computing fluid-dominated modes such as the fundamental $f$-mode and higher-order pressure ($p$-) modes, where the coupling to spacetime is relatively weak. While it does not capture spacetime-dominated modes (e.g., $w$-modes), it significantly reduces the complexity of the problem and retains good accuracy (typically within 20\%) for fluid modes, especially in compact stars. For instance, research on protoneutron stars has shown that while the approximation captures the qualitative behavior of oscillation frequencies, it can overestimate the $f$-mode frequency by approximately 20\% \cite{PhysRevD.102.063025}. Another study focusing on NSs with hyperonic matter found that the Cowling approximation can overestimate the quadrupolar $f$-mode frequency by up to 30\% compared to full general relativistic calculations \cite{PhysRevC.106.015805}.

In this work, we adopt the Cowling approximation to compute the non-radial oscillation modes of DM admixed NSs. The fluid perturbations are described by two radial functions, $W(r)$ and $V(r)$, which correspond to the radial and tangential/angular components of the Lagrangian fluid displacement, respectively. The perturbation equations under this approximation take the following form \cite{PhysRevD.99.123024}
%%%%%%%%%%%%%%%%%%%%%%%%%%%%%
\begin{eqnarray}
    \frac{dW}{dr} &=& \frac{1}{c_{s}^{2}} \left(\frac{d\Phi}{dr}W+\omega^{2}r^{2}e^{\lambda-2\Phi}V\right) - l(l+1)e^{\lambda}V, \\
    \frac{dV}{dr} &=& -\frac{1}{r^{2}}e^{\lambda}W + 2\frac{d\Phi}{dr}V-{\cal A}\left(\frac{1}{\omega^{2}r^{2}}\frac{d\Phi}{dr}e^{-\lambda+2\Phi}W+V\right), \nonumber \\
    \label{eq:radperteqs}
\end{eqnarray}
%%%%%%%%%%%%%%%%%%%%%%%%%%%%%
where $l$ denotes the azimuthal quantum number in spherical harmonic, $c_s^2$ is the adiabatic sound speed, defined as  $c_s^2 \equiv \left(\partial p/\partial\varepsilon\right)_s,$ and the quantity $\mathcal{A}(r)$, sometimes referred to as the Schwarzschild discriminant, is given by \begin{equation} \mathcal{A}(r) = \frac{1}{\varepsilon + p}\left(\varepsilon' - \frac{p'}{c_s^2}\right), \end{equation} with $\varepsilon' \equiv d\varepsilon/dr$ and $p' \equiv dp/dr$ evaluated from the equilibrium background configuration. The quantity $\mathcal{A}(r)$ determines the presence of buoyancy in the star and governs the existence of gravity ($g$-) modes, which arise in compositionally stratified stars. In such stars, $c_s^2$ differs from the equilibrium derivative $(dp/d\varepsilon)_{\rm eq}$, reflecting an adiabatic perturbation with different thermodynamic pathways. However, in this work, we assume a barotropic EOS and define the sound speed as $c_s^2 = (dp/d\varepsilon)_{\rm eq}$, implying no composition or entropy gradients. As a result, the Schwarzschild discriminant $\mathcal{A}(r)$ vanishes identically, and no $g$-modes are present. The system thus supports only the fundamental mode and its pressure-mode overtones. We emphasize that, in our model, the dark matter number density is not evolved dynamically but is prescribed as a function of the baryon density via Eq.~\eqref{eq:ndensity}. This leads to a barotropic composite equation of state where all thermodynamic quantities depend solely on the baryon density. As a result, the Schwarzschild discriminant $\mathcal{A}(r)$ vanishes identically, and no gravity ($g$-) modes appear in our analysis. However, we acknowledge that a more general treatment—where dark matter is treated as a dynamically independent fluid—could introduce composition gradients and associated $g$-modes.

To integrate this system of perturbation equations, one has to impose appropriate boundary conditions. Regularity conditions at the stellar center demand that the displacement functions behave as: 
\begin{equation} 
    W(r) = W_0\ r^{l+1} \hspace{0.6cm}  {\rm{and}}  \hspace{0.6cm} V(r) = -\frac{W_0}{l}\ r^{l+1}, 
\end{equation} 
which ensures a smooth, finite solution at $r = 0$ and $W_0$ is some arbitary constant. The outer boundary condition requires that the Lagrangian pressure perturbation vanishes at the surface of the star ($\Delta p = 0$), leading to: 
\begin{equation} 
    \frac{d\Phi}{dr}W + \omega^{2}r^{2}e^{\lambda-2\Phi}V = 0 \hspace{0.8cm} \text{at} \hspace{1cm} r = R. 
\end{equation}
%%%%%%%%%%%%%%%%%%%%%%%%%%%%%
Solving this boundary value problem yields the characteristic eigenfrequencies $\omega$ corresponding to the non-radial oscillation modes. In this study, we focus on the fundamental mode and the first pressure overtone ($p_{1}$-mode) for quadrupolar oscillations ($l=2$). These low-order fluid modes are the most relevant for gravitational wave observations and provide crucial insights into the internal structure of NSs and the impact of DM admixture.

Figure~\ref{fig:figure4} displays the variation of the non-radial $f$-mode frequency $f_{\rm f}$ as a function of stellar mass for NSs described by three RMF nuclear matter EOSs—IOPB-I, BigApple, and NL3—under different DM configurations. Each column corresponds to a different value of the DM steepness parameter: $\beta = 1, 2,$ and 4 (left to right). Solid curves represent purely baryonic stars, while dashed and dotted curves correspond to DM-admixed stars for a range of DM effective scaling parameter $\alpha M_{\chi}$, distinguished by different marker styles.

The $f$-mode arises from the fundamental non-radial fluid oscillation with no radial nodes, restored primarily a combination of pressure and gravitational forces, and is known to be strongly correlated with the average density of the star, typically scaling as $f_{\rm f} \propto \sqrt{M/R^{3}}$ \cite{PhysRevLett.77.4134, PhysRevD.99.123024}. This behavior is clearly observed from Fig.~\ref{fig:figureA1} in the Appendix. For a fixed EOS and steepness parameter $\beta$, increasing $\alpha M_{\chi}$ leads to higher $f_{\rm f}$ values, despite a reduction in total mass. This seemingly counterintuitive trend can be understood by noting that increasing DM content compresses the star more efficiently, leading to a higher average density—even as the mass decreases. The associated increase in $f_{\rm f}$ reflects the enhanced restoring force resulting from the more compact configuration.

Across the different EOSs, NL3 consistently yields the lowest $f_{\rm f}$ values at fixed mass. This is because, while NL3 is the stiffest EOS and supports the highest maximum masses, its stars have the largest radii for a given mass, resulting in the lowest average densities and hence lower $f_{\rm f}$. In contrast, BigApple and IOPB-I, being comparatively softer, produce more compact stars with higher average densities and thus larger $f_{\rm f}$.

An interesting crossing behavior is observed between the IOPB-I and BigApple curves. At low masses, BigApple predicts higher $f_{\rm f}$ values due to its ability to generate more compact stars in this regime. However, as the stellar mass increases, the IOPB-I configurations eventually become more compact than those of BigApple—leading to a crossover in the oscillation frequency behavior. This reflects the nontrivial interplay between the EOS stiffness and DM-induced structural modifications, especially in the high-mass end of the stable branch.

Finally, the impact of increasing $\beta$ is evident across columns in Fig.~\ref{fig:figure4}. Larger $\beta$ values result in steeper DM profiles, confining DM more tightly to the stellar core. This localized concentration enhances central pressure gradients, increasing the average density and leading to even higher $f_{\rm f}$ values for a given mass (For instance, in Fig. \ref{fig:figure4} if you focus on BigApple parameter set, $f_{\rm{f}}$ for a 1 $M_{\odot}$ stellar model with $\alpha M_{\chi} = 0.10$ and $\beta=4$ is $\approx 2.50$ kHz, while the same with $\beta = 2$ is $\approx 2.15$ kHz). The effect is particularly visible for high $\alpha M_{\chi}$, where the DM's gravitational influence is most pronounced.

The first pressure ($p_{1}$-) mode corresponds to the first overtone of pressure-driven non-radial oscillations and represents the next eigenmode above the fundamental mode. Unlike the $f$-mode, which lacks radial nodes, the $p_1$-mode contains one radial node and is governed more strongly by the local pressure gradients throughout the star. Figure~\ref{fig:figure5} displays the variation of the $p_1$-mode frequency, $f_{p_{1}}$, as a function of stellar mass for the same set of EOSs and DM configurations shown in Fig.~\ref{fig:figure4}. The column-wise structure and styling conventions (solid vs. dashed and dotted curves and marker styles) remain unchanged. As before, each sequence is truncated at the maximum mass configuration, corresponding to the onset of dynamical instability.

Unlike in the $f$-mode case, the $p_1$-mode curves for IOPB-I do not intersect those of BigApple for any $\beta$ or $\alpha M_{\chi}$ configuration. This behavior indicates that, across the entire stable mass range, the pressure-mode frequencies predicted by BigApple remain consistently higher than those of IOPB-I. The absence of crossing suggests that the relative ordering of the $p_1$-mode frequencies is more robust against variations in DM content, likely due to the stronger dependence of these modes on the overall pressure stratification rather than subtle changes in average density.

Another observation is that the variation in $f_{p_1}$ due to the inclusion of DM is generally less pronounced than for the $f$-mode, especially in the high $\beta$ regime. For $\beta = 4$, the $p_1$-mode  frequencies across different $\alpha M_{\chi}$ values remain remarkably close to the DM-free case, indicating a minimal sensitivity to the DM-induced modifications in the stellar structure. This contrasts with the $f$-mode, where the frequency shows a clearer upward shift with increasing $\alpha M_{\chi}$. The relative insensitivity of the $p_1$-mode at high $\beta$ likely stems from the sharply localized DM distribution near the core, which, while influencing the central compactness, contributes less significantly to the global pressure profile that governs higher-order pressure modes. However, this distinction becomes less clear at lower $\beta$. For $\beta = 1$, the variation in $f_{p_1}$ across DM configurations is somewhat comparable to that seen in the $f$-mode, reflecting the broader spatial extent of DM in these cases. The more distributed DM profile at low $\beta$ modifies the pressure gradients over a wider region, thereby exerting a more noticeable influence on both fundamental and overtone modes. 

This suggests that the $p_1$-mode is less sensitive to DM-induced changes in global structure, such as compactness or average density, and more tightly governed by the intrinsic stiffness and pressure gradients set by the nuclear EOS. This behavior highlights the interplay between DM localization and the modal sensitivity to different regions of the star, with $f$-modes being more responsive to global density and compactness, while $p_1$-modes depend more strongly on the detailed internal stratification.
%%%%%%%%%%%%%%%%%%%%%%%%%%%%%
%%%%%%%%%%%%%%%%%%%%%%%%%%%%%
\section{Testing Universal Relations with Dark Matter}
\label{sec:5}
%%%%%%%%%%%%%%%%%%%%%%%%%%%%%
%%%%%%%%%%%%%%%%%%%%%%%%%%%%%
% Figure 6
%%%%%%%%%%%%%%%%%%%%%%%%%%%%%
\begin{figure*}[htbp]
    \centering
    \includegraphics[width=\textwidth]{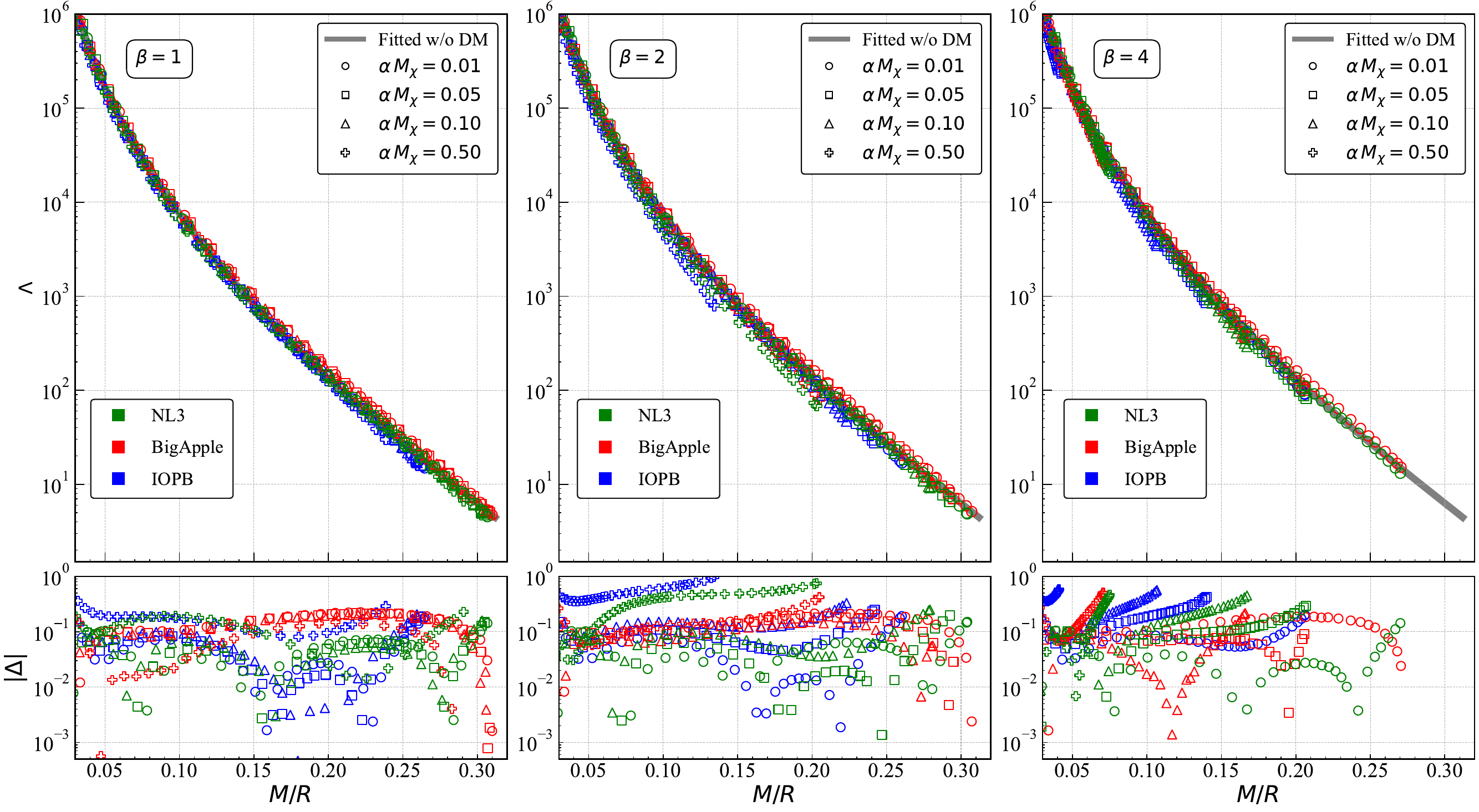}
    \caption{Dimensionless tidal deformability $\Lambda$ as a function of stellar compactness $C = M/R$ for three nuclear matter EOS models: IOPB-I (blue), BigApple (red), and NL3 (green), including DM-admixed stars for different values of the effective DM scaling parameter $\alpha M_{\chi} = 0.01$, 0.05, 0.10, and 0.50, indicated by distinct marker styles. Each panel corresponds to a fixed steepness parameter $\beta = 1$, 2, and 4 (left to right). The solid gray line represents the EOS-insensitive fit obtained from purely baryonic stars (without DM). The lower subpanels show the relative fractional deviation $\Delta = |\Lambda_{\rm actual} - \Lambda_{\rm fit}|/\Lambda_{\rm fit}$ from the fitted curve, quantifying the breakdown of the universal relation due to DM effects.}
    \label{fig:figure6}
\end{figure*}
%%%%%%%%%%%%%%%%%%%%%%%%%%%%%
%%%%%%%%%%%%%%%%%%%%%%%%%%%%%
% Figure 7
%%%%%%%%%%%%%%%%%%%%%%%%%%%%%
\begin{figure*}[htbp]
    \centering
    \includegraphics[width=\textwidth]{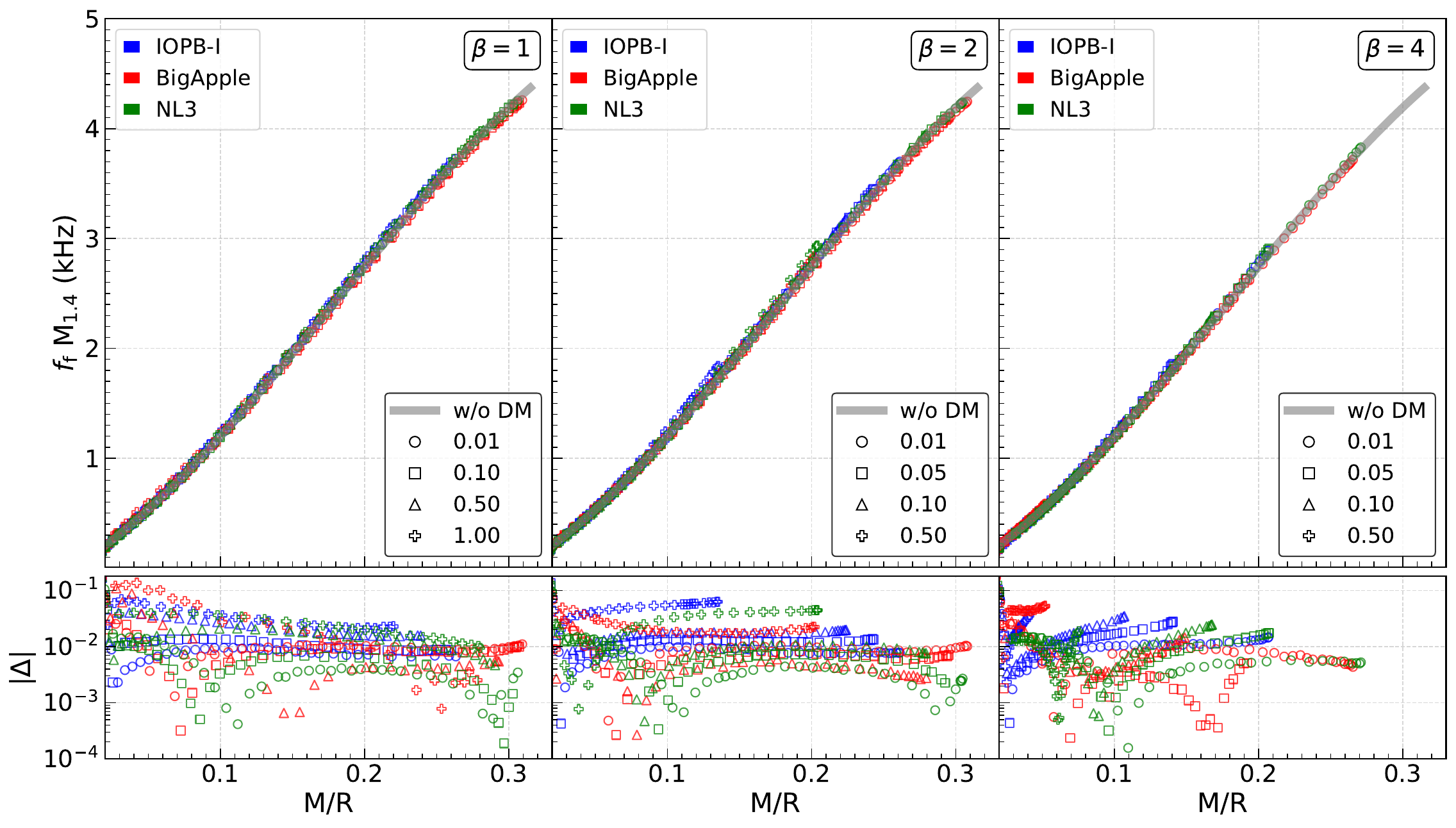}
    \caption{Same as Fig.~\ref{fig:figure6}, but for the universal relation between the mass-scaled fundamental mode frequency, $f_{\rm f}M_{1.4}$ (in kHz), and stellar compactness $C\equiv M/R$. Here, $M_{1.4}\equiv M/(1.4M_{\odot})$ is the stellar mass scaled by the canonical NS mass. The solid gray line represents the best-fit curve obtained from purely baryonic (i.e., non-DM-admixed) configurations, while the lower panels display the fractional deviation from this fit for different DM admixtures.}
    \label{fig:figure7}
\end{figure*}
%%%%%%%%%%%%%%%%%%%%%%%%%%%%%
Universal relations among macroscopic NS observables provide powerful tools for constraining stellar properties in a manner that is largely insensitive to the details of the underlying nuclear EOS. These relations connect (dimensionless) quantities such as the moment of inertia, tidal deformability, quadrupole moment, and $f$-mode frequencies, offering robust correlations that hold across a wide variety of EOS models. Their near-independence from microphysical uncertainties makes them especially valuable for interpreting astrophysical data—enabling indirect inference of difficult-to-measure quantities and serving as consistency checks in multimessenger observations.
%%%%%%%%%%%%%%%%%%%%%%%%%%%%%
% Figure 8
%%%%%%%%%%%%%%%%%%%%%%%%%%%%%
\begin{figure*}[htbp]
    \centering
    \includegraphics[width=\textwidth]{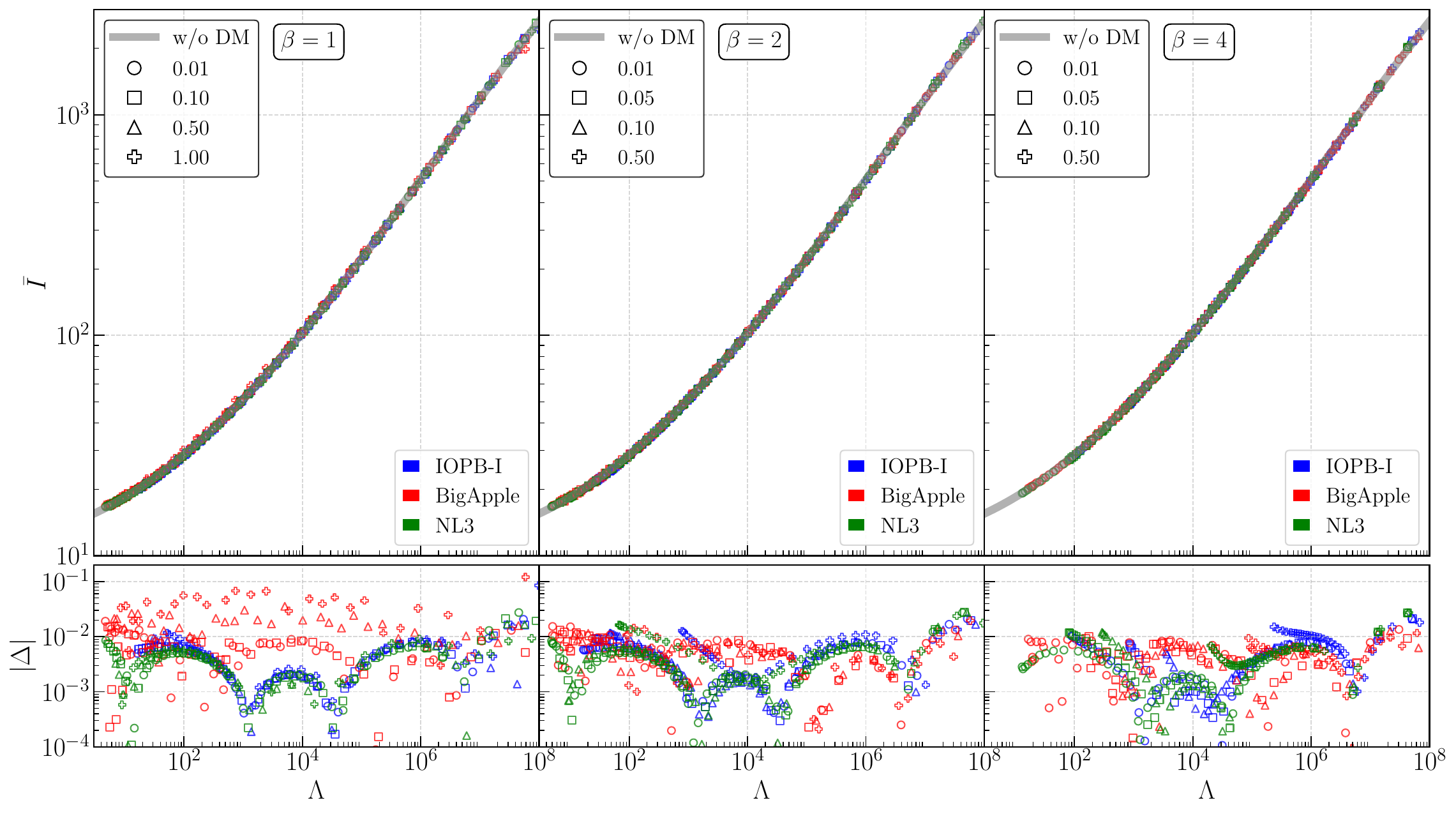}
    \caption{Same as Figs.~\ref{fig:figure6} and \ref{fig:figure7}, but for the dimensionless moment of inertia $\bar{I} \equiv I/M^{3}$ as a function of the dimensionless tidal deformability $\Lambda$ (i.e., the I-Love relation). The solid gray line represents the EOS-independent fit obtained from the configurations without DM, and the bottom panels show the fractional deviation $|\Delta| = |\bar{I}_{\rm actual} - \bar{I}_{\rm fit}|/\bar{I}_{\rm fit}$ for each EOS and DM configuration. Marker colors and styles follow the same conventions as in earlier figures.}
    \label{fig:figure8}
\end{figure*}
%%%%%%%%%%%%%%%%%%%%%%%%%%%%%

To explore whether the presence of DM alters or violates the established universal relations of NSs, we examine three widely studied correlations: (1) the tidal deformability and compactness relation ($\Lambda$-$C$) \cite{PhysRevD.88.023007}; (2) the mass-scaled $f$-mode frequency and compactness relation ($f_{\rm f} M$-$C$) \cite{10.1046/j.1365-8711.1998.01840.x, 10.1111/j.1365-2966.2005.08710.x}; and (3) the dimensionless moment of inertia and tidal deformability relation ($\bar{I}$-$\Lambda$), also known as the I-Love relation \cite{doi:10.1126/science.1236462}. These relations are known to hold with remarkable accuracy across a wide range of nuclear matter EOS models in the absence of exotic physics. Our aim is to assess the extent to which these relations are preserved or violated when DM is present in the NS interior.

For each relation, we begin by constructing a reference fit using NS configurations constructed purely from nuclear matter, i.e., without any DM contribution, for all three RMF EOSs considered in this work—IOPB-I, BigApple, and NL3. These fits provide EOS-independent baselines against which DM-admixed models are compared.

We then assess how the introduction of DM alters these relations by computing the relative deviation of DM-admixed configurations from the fitted baseline. For a given DM configuration (i.e., specified by $\alpha M_{\chi}$ and $\beta$), we evaluate the deviation in each observable relative to the baseline fit. Specifically, for the $\Lambda$-$C$ and $f_{\rm f} M$-$C$ relations, we define the relative deviation as \begin{equation} \Delta = \left| \frac{Y_{\rm actual} - Y_{\rm fit}}{Y_{\rm actual}} \right|, \label{eq:Delta}\end{equation} where $Y$ represents either $\Lambda$ or $f_{\rm f} M$ for each value of $C$, and ``actual" denotes the value computed from the DM-admixed model, while ``fit" denotes the value predicted by the fit from purely baryonic stars. Similarly, for the I-Love relation, the deviation is computed as 
\begin{equation} \Delta = \left| \frac{\bar{I}_{\rm actual} - \bar{I}_{\rm fit}}{\bar{I}_{\rm actual}} \right|, 
\label{eq:Delta1}\end{equation} 
with $\bar{I} = I/M^3$ for corresponding value of $\Lambda$.
In Figs.~\ref{fig:figure6}, \ref{fig:figure7}, and \ref{fig:figure8}, each universal relation is visualized along with its corresponding baseline fit (shown as a solid gray line), where the lower panels in each figure display the computed relative deviation $\Delta$ for various DM-admixed configurations. 

Figure~\ref{fig:figure6} displays the universal relation between the dimensionless tidal deformability $\Lambda$ \cite{PhysRevD.81.123016} and the stellar compactness $C\equiv M/R$, along with the corresponding relative deviations in the lower panels. The gray curve represents the baseline fit constructed from purely nuclear matter stars across the three RMF EOSs. The fitted relation takes the form:
\begin{equation}
    \log_{10} \Lambda =  \sum_{n=-1}^{3} a_{n} \left(\frac{5\ M}{R}\right)^{n},
    \label{eq:Lamda_MR_fitting}
\end{equation}
with coefficients $a_{-1} = 0.1641$, $a_{0} = 5.7791$, $a_{1} = 5.3095$, $a_{2} = 1.9191$ and $a_{3} = - 0.4275$.
We note that even among purely nuclear matter stars, the maximum deviation from this fitted relation reaches approximately 20\%, as quantified by the fractional deviation $\Delta_\Lambda^{\rm NM}$ defined in Eq.~(\ref{eq:Delta}). Considering this value of $\Delta_\Lambda^{\rm NM}$ as reference, our analysis shows that the $\Lambda$-$C$ relation remains largely intact across most of the DM parameter space. The DM-admixed configurations generally follow the same trend as the EOS-independent baseline fit, with deviations typically remaining modest. However, localized exceptions arise for specific combinations of steep DM profiles (e.g., $\beta = 2$) and high DM scaling ($\alpha M_{\chi} = 0.50$), where deviations near the maximum mass configuration can become large—approaching $\sim 100\%$ in fractional error. We emphasize that these cases are isolated outliers occurring at the edge of stability and do not reflect a systematic or global breakdown of the universal behavior. Additionally, for $\beta = 4$, a modest increase in deviation is again observed in stellar configurations at high $\alpha M_{\chi}$, suggesting a non-monotonic trend with respect to $\beta$ that may reflect the complex interplay between DM localization and stellar compactness. These outliers reflect the sensitivity of the $\Lambda$-$C$ relation to the DM distribution in extreme configurations, but the relation overall remains robust across the wide range of DM scenarios considered in this work.

Figure~\ref{fig:figure7} shows the universal relation between the mass-scaled fundamental mode frequency, i.e., $f_{{\rm f}} M_{1.4}$ and stellar compactness $M/R$, where $M_{1.4} \equiv M/(1.4 M_{\odot})$. The solid gray curve represents the baseline fit constructed from purely nuclear matter stars across the three RMF EOSs, described by the fitting function \cite{PhysRevD.105.023007}:
%%%%%%%%%%%%%%%%%%%%%%%%%%%%%
\begin{equation}
    f_{{\rm f}} M_{1.4} = \sum_{n=0}^{3}\ b_{n} \left(\frac{M}{R}\right)^{n} ,
    \label{eq:f-mode_MR_fitting}
\end{equation}
%%%%%%%%%%%%%%%%%%%%%%%%%%%%%
with coefficients $b_{0} = 0.0217$, $b_{1} = - 8.3413$, $b_{2} = 41.9344$ and $b_{3} = -77.9851$. 
We note that for purely nuclear matter stars, the maximum deviation from this fitted relation is approximately 4.5\%, as quantified by the fractional deviation $\Delta_{f_fM_{1.4}}^{\rm NM}$ using Eq.~(\ref{eq:Delta}). In contrast to the $\Lambda-C$ relation, the deviations in this case remain relatively moderate across the full range of DM parameters and EOSs considered. The relation is preserved with reasonable accuracy, especially for lower values of $\alpha M_{\chi}$. Even in cases with larger DM content and steep profiles, such as $\beta = 4$, the fractional deviations $\Delta$ remain contained, indicating that the scaling between $f_{\rm f}M_{1.4}$ and compactness is not strongly disrupted by the presence of DM. This result contrasts with earlier work in Ref.~\cite{PhysRevD.104.123006}, where a similar Higgs-portal DM model was considered using a constant DM Fermi momentum (i.e., uniform DM density) across the star. Under that assumption, the $f_{{\rm f}} M - C$ relation exhibited a much weaker universality, with notable dispersion among the curves for different DM parameters. In contrast, our physically motivated, variable DM density profile—tied to the baryon distribution—leads to more realistic spatial confinement of DM near the core, preserving the effective scaling behavior and yielding a significantly tighter adherence to the universal relation.
We note that another universal relation between the mass-scaled frequency and the tidal deformability is also known. We also check the validity of this type of universal relation in Appendix \ref{sec:appendx1}.

We now turn to the third and most robust correlation examined in this work: the I-Love relation connecting the dimensionless moment of inertia $\bar{I} \equiv I/M^{3}$ to the dimensionless tidal deformability $\Lambda$. To compute the moment of inertia $I$, we adopt the slow-rotation approximation developed by Hartle \cite{1967ApJ...150.1005H, 1968ApJ...153..807H}, wherein rotation is treated as a first-order perturbation on the static, spherically symmetric background. The resulting frame-dragging equation is solved using the equilibrium profiles obtained earlier, and the total moment of inertia is then evaluated via:
\begin{equation} 
    I = \frac{8\pi}{3} \int_0^R (\varepsilon + p) \left(1 - \frac{\omega(r)}{\Omega}\right)\ e^{\lambda-\Phi}\ r^4 dr. 
\end{equation} 
The result is normalized as $\bar{I} = I/M^{3}$, which serves as the dimensionless quantity relevant for the universal relation. Figure~\ref{fig:figure8} presents the $\bar{I}-\Lambda$ relation, along with the corresponding relative deviations in the lower panels. The gray curve denotes the EOS-insensitive baseline fit constructed from purely nuclear matter stellar configurations, given by the expression \cite{PhysRevD.88.023009, Landry_2018}: 
\begin{equation} 
    \log_{10} \bar{I} = \sum_{n=0}^{4} d_{n} \left(\log_{10}\Lambda\right)^{n}, \label{eq:ILove_fit} 
\end{equation} with coefficients 
$d_{0} = 1.1396$, $d_{1} = 0.0909$, $d_{2} = 0.0359$, $d_{3} = -6.8\times10^{-4}$ and $d_{4} = -9.6\times10^{-5}$. 
We note that for purely nuclear matter stars constructed using the considered RMF parameter sets, the maximum deviation from this fitted relation is approximately 2.5\%, as measured by the fractional deviation $\Delta_{\bar{I}}^{\rm NM}$ using Eq.~(\ref{eq:Delta1}).

The I-Love relation has previously been demonstrated to be one of the most robust universal relations in NS physics, and our results confirm that this holds even in the presence of DM. Across all DM-admixed configurations—spanning various values of $\alpha M_{\chi}$ and $\beta$—the deviations remain remarkably small, significantly smaller than those observed in the $\Lambda-C$ and $f_{\rm f} M_{1.4}-C$ relations. This exceptional robustness suggests that the I-Love relation is remarkably insensitive to the modifications induced by a nonuniform DM distribution. The tightly preserved trend across all configurations reinforces the potential of the I-Love relation as a reliable diagnostic tool for testing fundamental physics in NSs, even when exotic components like DM are present.
%%%%%%%%%%%%%%%%%%%%%%%%%%%%%
%%%%%%%%%%%%%%%%%%%%%%%%%%%%%
\section{Summary}
\label{sec:6}
%%%%%%%%%%%%%%%%%%%%%%%%%%%%%
%%%%%%%%%%%%%%%%%%%%%%%%%%%%%
In this work, we investigated the structural, dynamical, and oscillatory properties of DM admixed NSs within a single-fluid formalism, incorporating DM through an effective Higgs-portal interaction. Adopting three representative RMF nuclear matter EOSs—IOPB-I, BigApple, and NL3—we modeled the DM density using a physically motivated, baryon-density-dependent profile controlled by parameters $\alpha M_{\chi}$ and $\beta$, and analyzed its impact on equilibrium structure, stability, non-radial oscillations, and universal relations.

We first constructed background stellar configurations by solving the modified TOV equations and analyzed their stability via radial oscillations. Our results confirmed that the onset of dynamical instability, signaled by the vanishing of the fundamental radial mode frequency, coincides with the maximum mass point for each DM configuration. We found that increasing $\alpha M_{\chi}$ and $\beta$ leads to a lower maximum mass and altered compactness, with a nontrivial dependence of the characteristic fundamental frequency $f_{\xi}$ on DM distribution. Notably, configurations with sharply peaked DM profiles (high $\beta$) exhibit a reversal in frequency trends, revealing the importance of DM spatial support in shaping stellar stability.

Next, we computed the non-radial $f$-mode and first pressure ($p_1$-) mode frequencies using the relativistic Cowling approximation. The $f$-mode frequency was found to increase with DM content, reflecting enhanced average density from gravitational compression. The $p_1$-mode frequencies, in contrast, displayed more modest variation, highlighting their sensitivity to the detailed pressure stratification rather than global density. The behavior of oscillation modes varied across EOSs, with BigApple typically yielding higher frequencies due to its more compact configurations at high central densities.

Finally, we examined three widely studied EOS-insensitive universal relations—$\Lambda$-compactness, $f_{\rm f}M_{1.4}$-compactness, and the I-Love relation—and evaluated their robustness in the presence of DM. Our findings indicate that while the $\Lambda$-compactness and $f_{\rm f}M_{1.4}$-compactness relations exhibit localized deviations under certain extreme DM configurations
%are moderately affected under extreme DM configurations, 
they largely hold across the broader parameter space. The I-Love relation remains remarkably intact even for highly concentrated DM profiles, underscoring its robustness and continued applicability as a diagnostic tool in multimessenger astrophysics.

Overall, our study highlights the intricate ways in which DM influences NS structure, oscillation spectra, and macroscopic correlations, offering a consistent framework to probe the interplay between DM and nuclear matter in compact stars. Looking ahead, several extensions are worth pursuing. A natural next step involves adopting a two-fluid formalism to treat DM and baryons as distinct components, which would enable more realistic modeling of dynamical decoupling and differential oscillations between the two sectors. Additionally, incorporating rotational effects beyond the slow-rotation approximation, or exploring the impact of DM on magnetized NS, could reveal further observable signatures. These directions, in conjunction with multimessenger astrophysical observations, will be key to advancing our understanding of DM in the strong gravity regime.
%%%%%%%%%%%%%%%%%%%%%%%%%%%%%%%%%%%%%%%%%%%%%%%%
\acknowledgments
%%%%%%%%%%%%%%%%%%%%%%%%%%%%%%%%%%%%%%%%%%%%%%%%
This work is supported in part by Japan Society for the Promotion of Science (JSPS) KAKENHI Grant Numbers 
JP23K20848  % Kiban(B) by Sotani
and JP24KF0090. % by Sotani & Kumar
%%%%%%%%%%%%%%
\appendix
%%%%%%%%%%%%%%
%%%%%%%%%%%%%%
\section{Additional correlations between the $f$-mode frequency and stellar properties}
\label{sec:appendx1}
%%%%%%%%%%%%%% 
%%%%%%%%%%%%%%%%%%%%%%%%%%%%%
% Figure 9
%%%%%%%%%%%%%%%%%%%%%%%%%%%%%
\begin{figure*}[tbp]
    \centering
    \includegraphics[width=\textwidth]{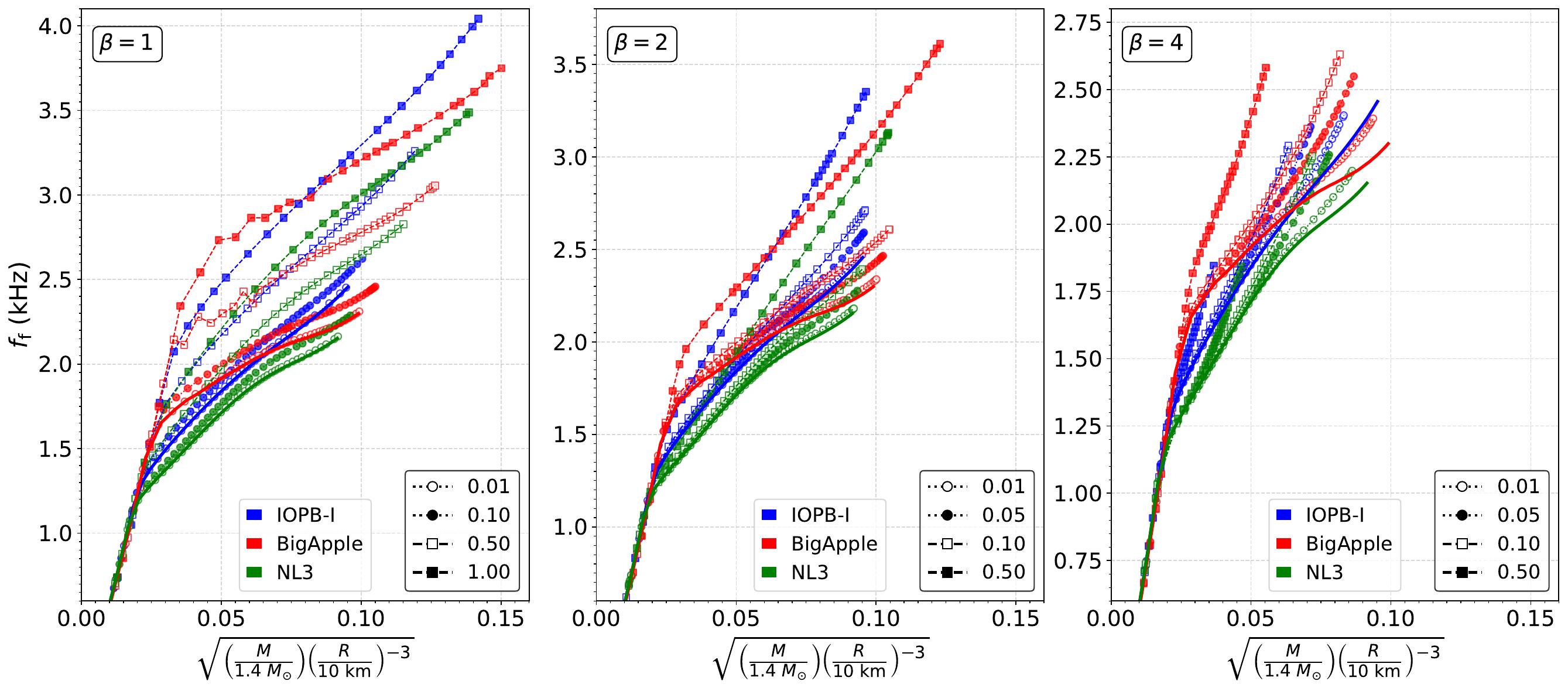} % Spans one column width
    \caption{Fundamental ($f$-mode) oscillation frequency $f_{\rm f}$ (in kHz) plotted as a function of the dimensionless average density parameter $x$ defined by Eq.~(\ref{eq:xx}) for DM-admixed NS configurations. Each panel corresponds to a fixed value of the DM density profile steepness parameter: $\beta = 1$ (left), $\beta = 2$ (middle), and $\beta = 4$ (right). Results are shown for three RMF nuclear matter EOSs: IOPB-I (blue), BigApple (red), and NL3 (green). Within each panel, curves are plotted for increasing values of the effective DM scaling parameter $\alpha M_{\chi}$, with numerical values labeled next to the plotting style in the legend box. For reference, the results for the stellar models without DM, i.e., purely composed of baryonic matter, are also shown with the solid lines. The frequency increases with average density, consistent with the expected scaling $f_{\rm f} \propto \sqrt{M/R^3}$, and is sensitive to both the nuclear matter EOS and the DM distribution.}
    \label{fig:figureA1}
\end{figure*}
%%%%%%%%%%%%%%%%%%%%%%%%%%%%%
%%%%%%%%%%%%%%%%%%%%%%%%%%%%%
% Figure 10
%%%%%%%%%%%%%%%%%%%%%%%%%%%%%
\begin{figure*}[tbp]
    \centering
    \includegraphics[width=\textwidth]{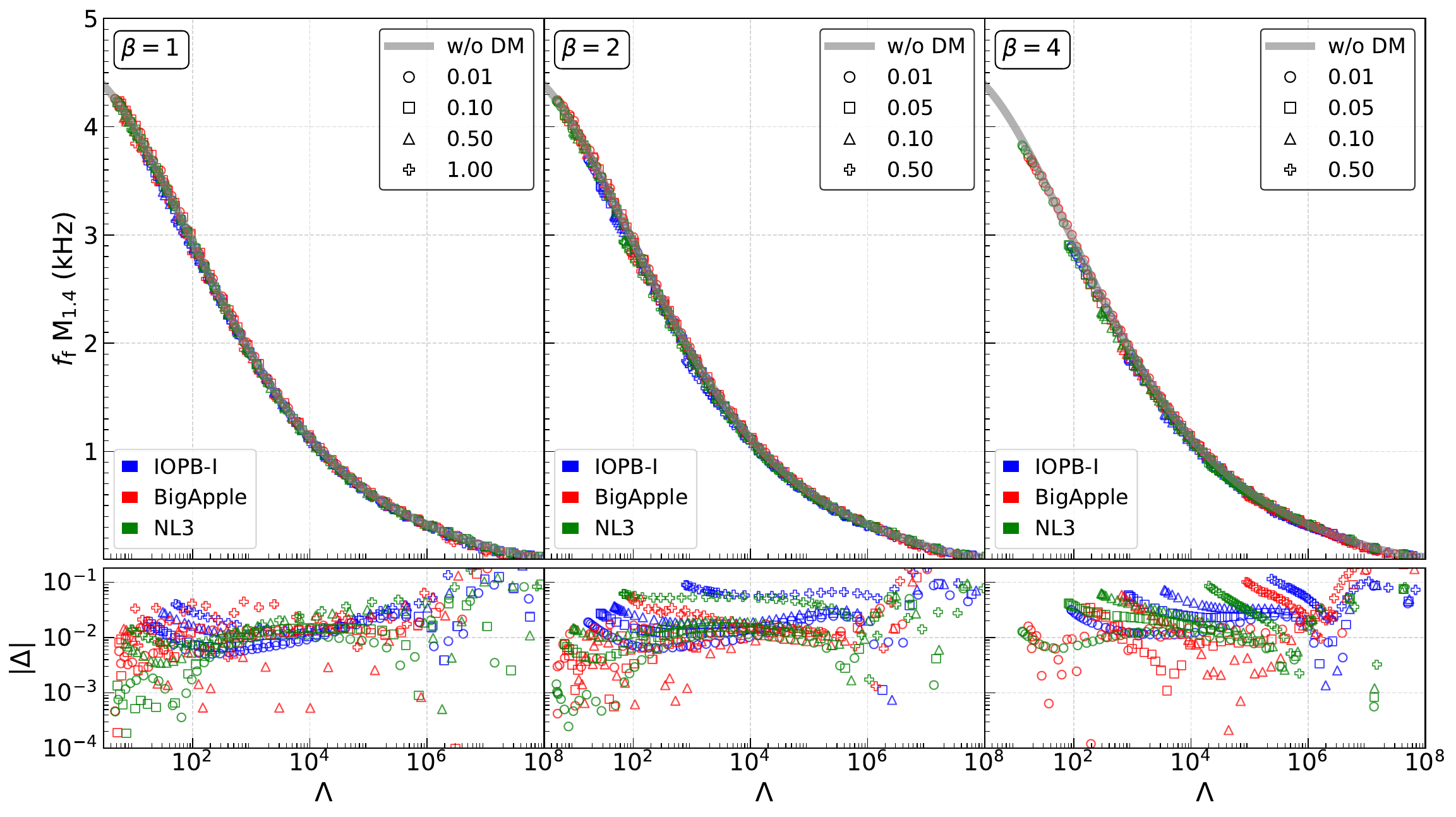} % Spans one column width
    \caption{Mass-scaled fundamental ($f$-mode) oscillation frequency $f_{\rm f} M_{1.4}$ (in kHz) plotted as a function of the dimensionless tidal deformability $\Lambda$ for DM-admixed NS configurations. Marker colors and styles follow the same conventions as in earlier figures (i.e. Fig. \ref{fig:figure7} and \ref{fig:figure8}). The solid gray curve represents the EOS-insensitive fit obtained from purely nuclear matter stars. Lower panels show the absolute fractional deviation $|\Delta|$ from the baseline fit, as defined by Eq. \eqref{eq:Delta}.}
    \label{fig:figureA2}
\end{figure*}
%%%%%%%%%%%%%%%%%%%%%%%%%%%%%
Originally, the correlation between the $f$-mode frequencies and stellar average density given by $\sqrt{M/R^3}$ has been pointed out in Refs.~\cite{PhysRevLett.77.4134, 10.1046/j.1365-8711.1998.01840.x}. Here, we explicitly verify this relation for DM admixed NSs by plotting the $f$-mode frequency $f_{\rm f}$ as a function of the dimensionless average density parameter,
\begin{equation}
 x\equiv \left(\frac{M}{1.4\ M_\odot}\right)\left(\frac{R}{10\ \mathrm{km}}\right)^{-3}. \label{eq:xx}
\end{equation}
The results, shown in Fig. \ref{fig:figureA1}, clearly demonstrate that $f_{\rm f}$ increases monotonically with the average density across all considered configurations. Each panel corresponds to a fixed value of the DM density profile steepness parameter $\beta = 1$, 2, and 4 (left to right), while different curves within each panel represent results for three RMF nuclear matter EOSs—IOPB-I, BigApple, and NL3—and for varying values of the effective DM scaling parameter $\alpha M_{\chi}$. Despite these variations, all sequences closely follow the same scaling trend, confirming the robustness of the $f_{\rm f} \propto \sqrt{M/R^3}$ behavior even in the presence of non-uniform dark matter. The modest variation in the curves among different EOSs and DM parameters reflects subleading effects, but the dominant dependence remains tied to the average density. This supports the interpretation of the $f$-mode as a global oscillation mode governed mainly by the star's average density, making it a useful probe for astrophysical observations.

In addition to the $f_{\rm f} M_{1.4}$-compactness correlation presented earlier in Fig.\ref{fig:figure7}, we explore here an alternative EOS-insensitive relation between the mass-scaled fundamental mode frequency and the dimensionless tidal deformability $\Lambda$. This correlation, previously discussed in~\cite{PhysRevD.104.123002}, connects the dynamical oscillation properties of the star to its tidal response. To examine this behavior in our DM-admixed NS models, we plot this relation in Fig. \ref{fig:figureA2}, where each panel corresponds to a fixed DM steepness parameter $\beta = 1$, 2, and 4 (from left to right). The plotted data include results from all three RMF EOSs (IOPB-I, BigApple, and NL3) for several values of the DM scaling parameter $\alpha M_{\chi}$. The reference fit (gray line) is constructed using purely baryonic configurations and takes the form:
\begin{equation}
    f_{{\rm f}} M_{1.4} = \sum_{n=0}^{5}\ k_{n} \left(\log_{10} \Lambda \right)^{n} ,
    \label{eq:f-mode_MR_fitting}
\end{equation}
where the best-fit coefficients obtained using purely nuclear matter stars constructed from the three considered RMF EOSs are $k_{0} = 4.523197$, $k_{1} = -0.003608$, $k_{2} = -0.705790$, $k_{3} = 0.193536$, $k_{4} = -0.020727$ and $k_{5} = 0.000809$. The upper panels of Fig.~\ref{fig:figureA2} show that all sequences, including DM-admixed cases, follow the same global trend, indicating that the $f_{\rm f} M_{1.4}$-$\Lambda$ relation remains valid even in the presence of DM. To quantify the accuracy of this relation, we compute the absolute fractional deviation $|\Delta|$ from the baseline fit using Eq. \eqref{eq:Delta}. For purely nuclear matter stars constructed from the considered RMF parameter sets, the maximum deviation remains within 2\% across the entire range of tidal deformabilities, confirming the robustness of the fitted relation. As seen in the lower panels of Fig. \ref{fig:figureA2}, the inclusion of DM induces mild to moderate deviations, with some configurations—particularly at higher values of $\Lambda$—exhibiting noticeable departures from the baseline. Nonetheless, the overall trend remains well preserved, and the $f_{\rm f} M_{1.4}$-$\Lambda$ correlation maintains a tighter universality compared to the corresponding $f_{\rm f} M_{1.4}$-compactness relation discussed earlier.
%%%%%%%%%%%%%%%%%%%%%%%%%%%%%
% References
%%%%%%%%%%%%%%%%%%%%%%%%%%%%%
\bibliographystyle{apsrev4-2}
\bibliography{main.bib} % Replace with your .bib file
\end{document}